\documentclass[sn-mathphys]{sn-jnl}

\newcommand{\AMSI}{\textsc{AMSI}}
\newcommand{\MuMFiM}{\textsc{MuMFiM}}
\newcommand{\FET}{$\text{FE}^2$}
\newcommand{\PUMI}{\textsc{PUMI}}

\usepackage[algo2e,linesnumbered]{algorithm2e}
\usepackage{physics}
\usepackage{subcaption}
\graphicspath{{figures/}}
\usepackage{xcolor}
\begin{document}
\lstset{
    language=C++,
    frame=single,
    numbers=left,
}

\renewcommand{\textcolor}[1]{}

\title[Open Source Framework For Fibrous Materials]{A new open source framework for multiscale modeling of fibrous materials on heterogeneous supercomputers}

\author[1]{\fnm{Jacob } \sur{Merson}}\email{mersoj2@rpi.edu}
\author[1]{\fnm{Catalin} \sur{Picu}}\email{picuc@rpi.edu}
\author[1]{\fnm{Mark S.} \sur{Shephard}}\email{shephard@rpi.edu}

\affil[1]{\orgdiv{Department of Mechanical Aerospace and Nuclear Engineering}, \orgname{Rensselaer Polytechnic Institute}, \orgaddress{\street{110 8th St.}, \city{Troy}, \state{NY} \postcode{12180} \country{USA}}}

\abstract{This article presents \MuMFiM{}, an open source application for multiscale modeling of fibrous materials on massively parallel computers. \MuMFiM{} uses two scales to represent fibrous materials such as biological network materials (extracellular matrix, connective tissue, etc.). It is designed to make use of multiple levels of parallelism, including distributed parallelism of the macro and microscales as well as GPU accelerated data-parallelism of the microscale. Scaling results of the GPU accelerated microscale show that solving microscale problems concurrently on the GPU can lead to a 1000x speedup over the solution of a single RVE on the GPU. In addition, we show nearly optimal strong and weak scaling results of \MuMFiM{} on up to 128 nodes of AiMOS (Rensselaer Polytechnic Institute) which is composed of IBM AC922 nodes with 6 Volta V100 GPU and 2 20 core Power 9 CPUs each. We also show how \MuMFiM{} can be used to solve problems of interest to the broader engineering community, in particular providing an example of the facet capsule ligament (FCL) of the human spine undergoing uniaxial extension.}

\maketitle

\section{Introduction}
Computing the response of materials under various loading conditions is at the heart of many important engineering questions. In some cases, the underlying material constitutive response is unknown, or, it is important to capture the micromechanical response of discrete constituents such as fibers, molecules, and so forth \cite{kanouteMultiscaleMethodsComposites2009}. In these cases the analyst can turn to multiscale methods. Broadly, multiscale methods are split into partitioned domain and hierarchical methods. Partitioned domain methods use different physical descriptions in different domains to capture behaviors at multiple length scales. These methods are useful when there is a well defined domain where fine scale behavior is occurring and they do not require strong scale separation. Alternatively, hierarchical methods include behavior at all length scales throughout the full domain and require strong scale separation. In hierarchical methods, the macroscale material properties are constructed through homogenization of the microscale material properties. This homogenization can be done a priori or concurrently with the macroscale solution procedure.

In this work, we use a hierarchical multiscale method where the homogenization is done concurrently with the macroscale solution procedure. This enables tracking of micromechanical properties, such as fiber orientations, which have functional significance.

The use of the hierarchical multiscale approach on two scales using finite elements at each scale was first proposed by Renard in 1987 and later generalized and implemented by Feyel in 1998 to study composite materials undergoing small strains \cite{kanouteMultiscaleMethodsComposites2009}. Feyel later coined the name \FET\ in reference to the use of finite elements at two scales and this name has received widespread adoption \cite{feyelMultiscaleFE2Elastoviscoplastic1999,feyelMultiscaleNonLinear2001,nezamabadiMultilevelComputationalStrategy2009,tikarrouchineThreedimensionalFE2Method2018,eidelHeterogeneousMultiscaleFinite2018}. The \FET\ method has been extended to large strains and nonlocal continua, however, it has been primarily limited to small problem sizes on the order of a few thousand elements due to the computational cost \cite{mieheComputationalMicrotomacroTransitions2003,kanouteMultiscaleMethodsComposites2009,smitPredictionMechanicalBehavior1998,feyelMultiscaleNonlinearFE22001,feyelMultilevelFiniteElement2003}. 

\textcolor{red} {
Material homogenization schemes have a long history in the computational mechanics community. They have been particularly successful in modeling composite materials with complex microstructures and nonlinear material behaviors \cite{castanedaEffectiveMechanicalProperties1991,moulinecFastNumericalMethod1994,moulinecNumericalMethodComputing1998,michelEffectivePropertiesComposite1999,geersHomogenizationMethodsMultiscale2017}. Over the past 20 years, considerable attention has been paid to incorporating history dependent behavior, localization, high order and micromorphic continuum descriptions and even applying numerical homogenization schemes to non-Newtonian fluids \cite{kouznetsovaApproachMicromacroModeling2001,coenenMultiscaleApproachBridge2012,pontecastanedaVariationalLinearComparison2023,kouznetsovaMultiscaleSecondorderComputational2004,temizerNumericalMethodHomogenization2007,zhiDirectFEModeling2022}. Another key aspect of development has been new methodologies that can reduce the computational cost of numerical homogenization such as \cite{temizerComputationMacroscopicTangent2008} which uses a penalty method to reduce the cost of the condensation method for computing the tangent material stiffness.
}

This paper introduces a general purpose large-strain \FET\ code, \MuMFiM\, that is designed to run on leadership class computing resources.\footnote{\MuMFiM\ can be found at \url{https://github.com/SCOREC/mumfim}.} \MuMFiM{} uses two levels of parallelism to achieve adequate performance across the two physical scales: The first level of parallelism is on the macroscale; we make use of the parallel unstructured mesh infrastructure (\PUMI{}) \cite{ibanezPUMIParallelUnstructured2016} to support distributed meshes and PETSc \cite{balayEfficientManagementParallelism1997,balayPETScUsersManual2018} to solve the resulting systems of linear equations. The second level of parallelism employs data parallel algorithms for the simultaneous solution of multiple microscale problems on GPUs.

\MuMFiM{} uses an updated-Lagrangian method to deal with large strains on the macroscale. This is important because the application of our code to biological tissues often requires strains in excess of 50\%. Our design is modular, which allows for the easy addition of new physics on both the macro- and microscales. When using our multiscale framework, the macroscale solver only requires a constitutive response function---namely, the microscale solver---that can compute the stress and material stiffness tensor for an increment in the deformation gradient. Due to the nonlinearities associated with the microscale fiber networks, a dynamic relaxation method \cite{underwoodDynamicRelaxation1986} has been employed to obtain the static response on the microscale.

This work contains the following contributions:
\begin{itemize}
    \item Open source two-scale finite element code (\MuMFiM) designed to run on massively parallel computers with GPU accelerators.
    \item Novel microscale solution algorithm, enabling the solution of highly nonlinear problems at the microscale.
    \item New method for computing the tangent material stiffness.
    \item Novel parallelization strategy for the microscale finite element computation.
    \item Demonstration of the parallel performance of \MuMFiM.
    \item Demonstration of \MuMFiM\ on a problem of biological importance.
\end{itemize}

\MuMFiM{} extends the state of the art of \FET{} both by increasing the scale of problems solved through heterogeneous supercomputers, and by developing novel methodologies that enable the inclusion of unstabilized stochastic fiber networks as a microscale material.

\textcolor{red}{Our framework shares the same goal as \cite{mahutgaNonaffineFiberNetwork2023}; to provide an open source multiscale simulation framework for fibrous materials. However, there are philosophical differences that drive a divergent approach. In particular, our previous work has shown that obtaining size-effect converged stiffness measurements in Voronoi fiber networks requires model sizes that are $\gg 40$ times the mean fiber length. Solving networks of this size is extremely expensive and requires the types of specialized GPU solvers that are employed here. In our experience, solving the full range of network deformations is not possible with gradient based nonlinear solvers such as Newton-Raphson that are used elsewhere \cite{mahutgaNonaffineFiberNetwork2023}. To this end, we use dynamic relaxation methods that do not require an invertable tangent stiffness matrix.}

This article is organized as follows: Section \ref{sec:multiscale_problem_formulation} gives a mathematical overview of the multscale framework employed here.
Section \ref{sec:finite_element_implementation} provides the finite element implementation and solution methodologies employed at the macro and micro scales.
Section \ref{sec:software_arch} describes the software components and parallelization strategy in \MuMFiM{}.
Section \ref{sec:PerformanceResults} provides performance results for our GPU accelerated microscale solver and the strong and weak scaling of \MuMFiM{} as a complete distributed parallel application.
Section \ref{sec:ExFCL} demonstrates the use of \MuMFiM{} to solve a problem of biological interest.
And, section \ref{sec:Conclusion} provides the concluding remarks and likely directions for future enhancements to \MuMFiM{}.

\section{Multiscale Problem Formulation}\label{sec:multiscale_problem_formulation}
\MuMFiM{} employs a hierarchical multiscale method which consists of a Cauchy continuum at the macroscale. At each macroscale material point \(\vb{X}^\text{M}\), the constitutive response function is computed from an embedded material that we call the microscale.
The multiscale formulation starts with the classic definition of the macroscale virtual power for a system in static equilibrium with no body forces
\begin{equation}
   \delta P = \delta P^\text{int}-\delta P^\text{ext} = 0 \quad \forall \delta \vb{v} \in \mathscr{W} \label{eq:virtual-power}
\end{equation}
where 
\begin{align}
    \delta P^\text{int} &= \int_V \sigma_{ij} \delta D_{ij} dV, \label{eq:virtual-internal-power} \\
    \delta P^\text{ext} &= \int_{\partial V} t_i \delta v_i d\Gamma, \label{eq:virtual-external-power}
\end{align}
\(\mathscr{W}\) is the set of kinematically admissible velocity fields, \(\sigma_{ij}\) is the macroscopic Cauchy stress, \(D_{ij}\) is the symmetric part of the velocity gradient, \(t_i\) is the surface tractions, \(V\) is the total body volume, and \(\partial V\) is the boundary to the body.

The microscale coordinates are written as a Taylor series expansion about the macroscale material point taken where the centroid of the microscale material element (RVE) is coincident with the material point. That is,
\begin{equation}
    x_i^{\text{m}}\left(\vb{X}^{\text{M}},\vb{X}^{\text{m}} \right) = X_i^{\text{M}} +  F_{ij}^{\text{M}} X_j^{\text{m}} + h_i\left( \vb{X}^{\text{m}}\right),
    \label{eq:micro_coords}
\end{equation}
where \(h_i\) is the higher order microscale fluctuations. The macroscale and microscale quantities have an majuscule or minuscule m superscript respectively. Differentiating equation \eqref{eq:micro_coords} gives the microscale deformation gradient
\begin{equation}
    F_{ij}^{\text{m}} = F_{ij}^{\text{M}} + \pdv{h_i}{X_j^\text{m}}.
    \label{eq:micro_fluctuations}
\end{equation}
This leads to the kinematically admissible microscale virtual velocity field
\begin{equation}
    \delta v_i^{\text{m}} = \delta \dot{F}_{ij}^{\text{M}} X_j^{\text{m}} + \delta \dot{h}_i,
\end{equation}
and virtual velocity gradient
\begin{equation}
    \delta D_{ij}^{\text{m}} = \text{sym} \left( \delta \dot{F}_{ik}^\text{M} \left(F_{kj}^\text{m}\right)^{-1} + \frac{\partial \delta \dot{h}_i}{\partial x^m_j} \right).
\end{equation}

Hill and others postulate that the mean fluctuation field is zero and thus
\begin{equation}
    F_{ij}^{\text{M}} = \int_{V^\text{m}}F_{ij}^{\text{m}} dV = \left\langle F_{ij}^{\text{m}} \right\rangle.
    \label{eq:homogenized_deformation_gradient}
\end{equation}
\cite{glugeGeneralizedBoundaryConditions2013,hillElasticPropertiesReinforced1963,liuDiscreteAveragingRelations2016}. Luscher and coauthors showed that this can also be interpreted as a requirement of orthogonality in mean and fluctuating fields \cite{walters2021considering,luscher2012essential}. Here, \(V^\text{m}\) corresponds to the volume of the microscale RVE. Simply stated, the deformation gradient at the macroscale material point is the volume average of the microscale deformation gradient. It is constructive to emphasize that the microscale RVE is assumed to be a differential volume compared with the macroscopic quantities and thus strong scale separation is needed. In other words,
\begin{equation}
    V > \sum V^m.
\end{equation}

The Hill-Mandel principle states that the variation of the volume average of the microscale power should be equal to the variation of macroscale power at the material point. That is,
\begin{equation}
    \frac{1}{V^\text{m}} \int_{V^\text{m}} \sigma_{ij}^{\text{m}} \delta D_{ij}^{\text{m}} dV = \sigma_{ij}^{\text{M}} \delta D_{ij}^{\text{M}}.
    \label{eq:hill-mandel}
\end{equation}
As a consequence of equation \eqref{eq:hill-mandel}, we are able to identify the macroscale Cauchy stress as the volume average of the microscale Cauchy stress for a system in static equilibrium with no body forces. Namely,
\begin{equation}
    \sigma_{ij}^{\text{M}} = \frac{1}{V^\text{m}} \int_{V^\text{m}} \sigma_{ij}^{\text{m}} dV = 
    \frac{1}{V^\text{m}} \int_{\partial V^\text{m}} t_i x_j^{\text{m}} d\Gamma.
\end{equation}
and restate the Hill-Mandel principle as
\begin{equation}
    \int_{\partial V^\text{m}} t_i \cdot \delta \dot{h}_i d \Gamma = 0.
    \label{eq:hill-mandel-2}
\end{equation}

In this work we use a homogeneous displacement boundary condition on the microscale, typically called the affine boundary condition, which states that
\begin{equation}
    x_i^{\text{m}} = X_i^{\text{M}} + F_{ij}^{\text{M}} X_j^{\text{m}},\quad \forall X_i^{\text{m}  } \in \partial V^\text{m}.
    \label{eq:affine-bc}
\end{equation}
By comparing equation \eqref{eq:affine-bc} to equation \eqref{eq:micro_coords} we see that \(h_i\) is identically zero on the boundary, and therefore this boundary condition trivially satisfies equation \eqref{eq:hill-mandel-2}, ergo the Hill-Mandel principle.

This choice is not unique. In fact, \cite{mersonSizeEffectsRandom2020} showed that the generalized boundary conditions---originally developed by \cite{glugeGeneralizedBoundaryConditions2013} for continuum applications---can be applied to fibrous materials to reduce the required RVE size, and will likely be included in future versions of \MuMFiM{}. For periodic materials, another common choice is to use periodic boundary conditions \cite{liuDiscreteAveragingRelations2016}. Generally, periodic boundary conditions should not be used for modeling fibrous materials, as they do not tend to have periodic structures.

\section{Finite Element Implementation} \label{sec:finite_element_implementation}
\MuMFiM\ makes use of the \FET\ two-scale finite element method. In this method, the macroscale is represented by a traditional finite element discretization. At each integration point of the macroscale, there is a microscale problem that is solved to obtain the constitutive response of the homogenized materials, that is, the stress and the fourth order material stiffness tensor.

 The homogenization scheme described in section \ref{sec:multiscale_problem_formulation} relies on two fundamental assumptions: first, the physical dimensions of the microscale must be at least one order of magnitude smaller than the encompassing mesh element. Second, the microscale must be large enough to exhibit representative behavior.
 
 In the context of solid mechanics problems, the material response can be considered representative when it is invariant to equivalent boundary conditions (e.g., equivalent homogeneous traction and homogeneous displacement). In the homogenization literature, a volume that exhibits this behavior is typically called a representative volume element (RVE). The required RVE size for fiber networks has been explored in detail in both 2D and 3D \cite{mersonSizeEffectsRandom2020,shahsavariSizeEffectMechanical2013}.
 
\subsection{Macroscale Finite Element Method} \label{sec:macro-finite-element}
 On the macroscale, \MuMFiM{} uses an updated-Lagrangian Galerkin finite element method with the weak form given by equations \eqref{eq:virtual-power}--\eqref{eq:virtual-external-power}. A brief description is included here for completeness.

Discretizing equation \eqref{eq:virtual-power}--\eqref{eq:virtual-external-power} gives
\begin{equation}
    \delta P = \delta v_{kI} \underbrace{\int_V B_{ijkI} \sigma_{ij}  dV}_{\vb{f}^\text{int}}- \delta v_{kI} \underbrace{\int_{\partial V} N_{I} t_{k} d\Gamma}_{\vb{f}^\text{ext}}=0.
    \label{eq:residual}
\end{equation}
where, majuscule subscripts refer to nodal indexes and miniscule subscripts refer to spatial indexes, \(B_{ijkI}\) is the symmetric gradient of the shape functions at node \(I\) (commonly called the strain displacement matrix) given by
\begin{equation}
    B_{ijkI} = \frac{1}{2}\left(\delta_{ki} \pdv{N_I}{x_j}+\delta_{kj} \pdv{N_I}{x_i} \right)
\end{equation}
and \(N_I\) is the shape functions.

By invoking the arbitrariness of the variation in nodal velocities we obtain the typical force residual equation
\begin{equation}
    \mathcal{R}_{kI} = \int_V B_{ijkI} \sigma_{ij}  dV- \int_{\partial V_j} N_{I} t_{k} d\Gamma_j = f_{kI}^\text{int}-f_{kI}^{ext} = 0.
    \label{eq:force-residual}
\end{equation}

A Newton-Raphson scheme is used to minimize the force residual which requires the linearization of the force residual with the displacements. The linearization of the internal force is included to emphasize the need for computing the tangent material stiffness in the multiscale framework. The material and geometric parts of the element tangent stiffness for the internal forces \(\frac{\partial f_iI^\text{int}}{\partial u_{jJ\ \ }}\) is given by
\begin{equation}
    K_{kIrJ}= \frac{\partial f^\text{int}_{kI}}{\partial u_{rJ\ }} = \int_V B_{ijkI} C_{ijpq} B_{pq rJ} dV+
    \delta_{ij}\int_V B_{ijkI} \sigma_{pq} B_{pqrJ} dV.
    \label{eq:tangent-stiffness}
\end{equation}
In practice, there is also a tangent stiffness contribution from the external forces when they transform with the displacements. This contribution has been omitted here for brevity. Details on computing the microscale material stiffness are given in section \ref{sec:microscale_solution_procedure}.

\begin{figure}
    \centering
    \includegraphics[width=0.6\textwidth]{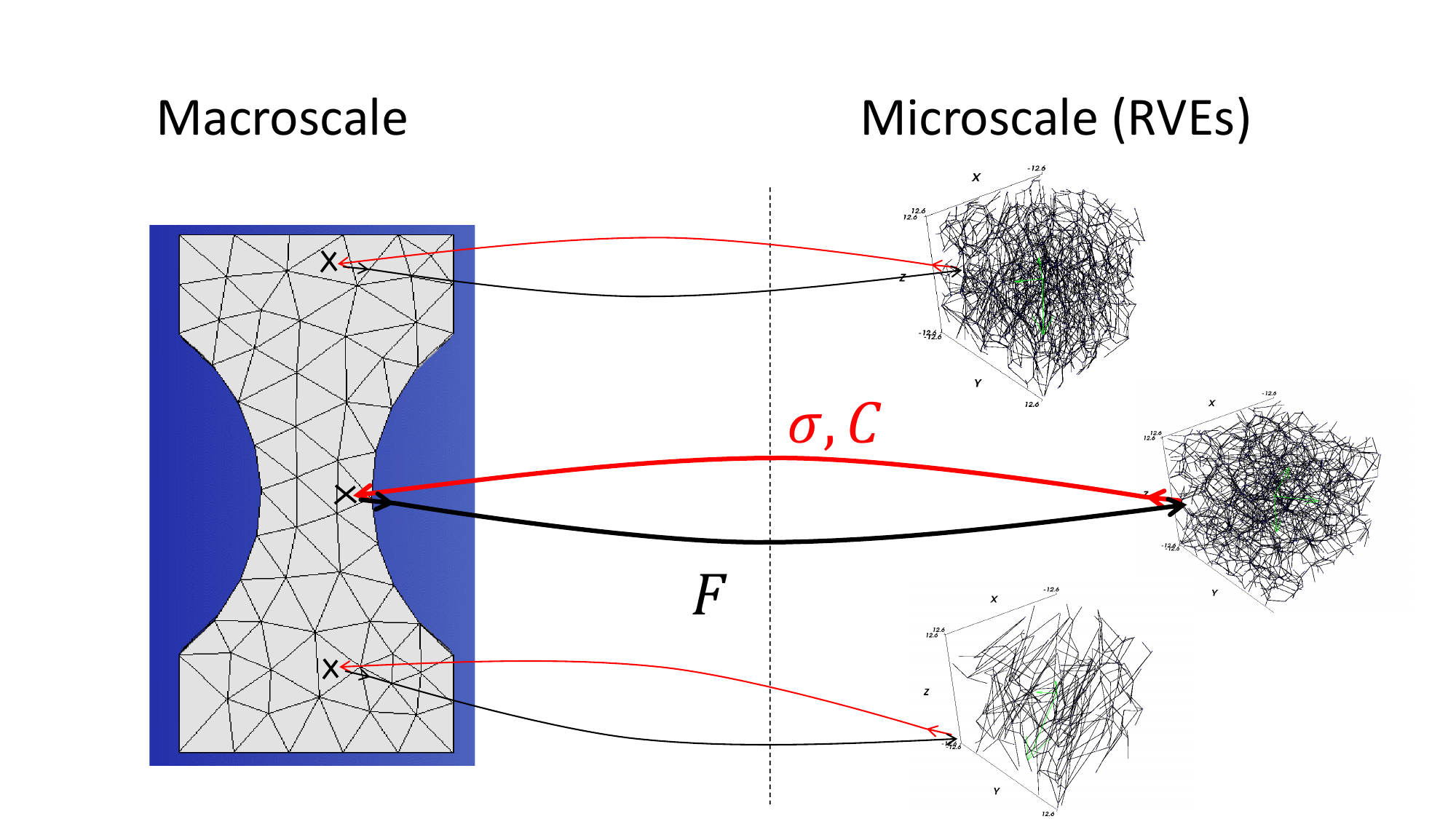}
    \caption{Schematic representation of the \FET\ multiscale finite element scheme used by \MuMFiM\ where material properties at each integration point are computed by a microscale finite element problem.}
    \label{fig:coupling_diagram}
\end{figure}

\begin{algorithm2e}
\KwData{macroscale mesh, microscale meshes, macroscale boundary conditions}
\KwResult{macroscopic displacements, stress}
Load macroscale and microscale meshes\;
Compute $\vb*{\sigma}$ and $\vb{C}$ from continuum estimate\;\label{alg:macroFET:continuum_estimate}
\While{not converged}
{
Compute increment in displacement $\delta \vb{u}=\vb{K}^{-1}\vb{R}(\vb{u})$\;
$\vb{u}^{n+1}\leftarrow \vb{u}^n+\delta \vb{u}$\;
Compute increment in deformation gradient, $\delta \vb{F}$, from $\delta \vb{u}$\;
Send $\delta \vb{F}$, to the microscale\;\label{alg:macroFET:send}
\nl Compute $\vb*{\sigma}$ and $\vb{A}=\frac{\partial \vb*{S}}{\partial \vb{E}}$ from microscale RVE\;\label{alg:macroFET:microsol}
Push forward $\vb{A}$ to get spatial stiffness $\vb{C}$\;\label{alg:macroFET:microsol2}
Send $\vb*{\sigma}$ and $\vb{C}$ to the macroscale\;\label{alg:macroFET:receive}
Compute macroscale convergence criteria\;
}
Write simulation data to file\;
\caption{Macroscale perspective of the \FET{} method used in \MuMFiM{}.} \label{alg:macroFET}
\end{algorithm2e}

Algorithm \ref{alg:macroFET} gives the macroscopic perspective of the \FET\ method used in \MuMFiM{}. Importantly, this shows how the macroscale solution proceeds identically to a standard Newton-Raphson scheme with three major exceptions. First, after computing a Newton step, the increment in the deformation gradient must be sent from the macroscale to the microscale (line \ref{alg:macroFET:send}). Second, the stress and material stiffness are computed by a secondary solver, not a analytic equation (lines \ref{alg:macroFET:microsol}--\ref{alg:macroFET:microsol2}). Lastly, the macroscale solver must receive the stress and material stiffness from the independent microscale solver (line \ref{alg:macroFET:receive}).

\subsection{Microscale Solution Procedure} \label{sec:microscale_solution_procedure}
In this work, the microscale is implemented as a large displacement truss model which can be used to model collagen fiber networks. With the fiber model, the solution of a partial differential equation is required to obtain the stresses for a given level of strain. We implement two methods: implicit and dynamic relaxation. The implicit method is a standard updated-Lagrangian method equivalent to the one described in section \ref{sec:macro-finite-element}.

We will restrict our discussion to the dynamic relaxation method because the response of fiber networks undergoing deformation is highly nonlinear and implicit methods are typically unable to achieve convergence. Networks go through local bifurcation points, or may be sub-isostatic; that is, the tangent stiffness matrix can be singular during an analysis \cite{licupElasticRegimesSubisostatic2016}. As a result, athermal fiber networks, such as collagen networks, are typically modeled with explicit methods \cite{islamStochasticContinuumModel2018,deogekarStrengthRandomFiber2018}. In previous works related to quasi-static loading of fibrous materials, authors have argued that if the ratio of kinetic energy to total energy is less than 5\%, then the resulting stress, strain, and material stiffness match the quasi-static case \cite{deogekarStrengthRandomFiber2018}. However, the presence of kinetic energy in the microscale is a violation of the Hill-Mandel principle for a static multiscale system. From a pragmatic perspective, the loss of kinetic energy in the micro-to-macro transition prevents convergence of the macroscale solver.

To work around these issues, a dynamic relaxation method is used for the microscale problems. Dynamic relaxation works by mapping a static analysis to a damped dynamic explicit analysis where the system residual is monitored for convergence \cite{underwoodDynamicRelaxation1986}. This method ensures that the kinetic energy is reduced to a desired tolerance.

The use of the dynamic relaxation method at the microscale increases the strain levels which the multiscale method can converge. However, it is subject to many of the same trade-offs as other explicit methods; that is, it does not require the formation or inversion of a global stiffness matrix but has a stable time step limited by the smallest element size via the Courant-Friedrichs-Lewey (CFL) condition \cite{courantUberPartiellenDifferenzengleichungen1928}.

\begin{algorithm2e}[!htp]
\SetKwFunction{GIF}{getInternalForces}
\SetKwProg{Fn}{Function}{ is}{end}
\Fn{\GIF}{
\tcc{Computes the internal forces at each node}
    \ForEach{fiber}
    {
        compute the fiber length\;
        compute the fiber contribution to the nodal force\;
        scatter fiber forces to nodes\;
    }
}
Load RVE mesh and compute edge connectivity\;
Compute the RVE mesh connectivity array\;
Compute diagonal mass matrix ($m$) \tcp*{finite element integration}
Transfer the connectivity array and mass matrix to the device (GPU)\;
Set displacement boundary conditions on fixed nodes \tcp*{fixed dof vector operation}
$\vb{v} \gets \vb{v_0}$, $\vb{u(x)}\vert_{\vb{x} \in \Gamma} \gets \vb{u_0}$, $\vb{u(x)}\vert_{\vb{x} \notin \Gamma}  \gets \vb{0}$, $n\gets 1$,$\vb{R} \gets \norm{\vb{1}}$,$\epsilon \gets 1\times 10^{-6}$\;
Get internal forces from fiber constitutive equation\;
 $\vb{f}^{0,\text{int}}\gets$ \GIF{$\vb{u}$}\;
Compute damping force \tcp*{free dof vector operation}
$\vb{f}^{0,\text{damp}} \gets c\vb{m}\vb{v}$\;

 $\vb{a^0} \gets 1/\vb{m}\left(\vb{f}^{0,\text{ext}}-\vb{f}^{0,\text{int}}-\vb{f}^{0,\text{damp}}\right)$\;
 $R\gets \norm{\vb{f}^{0,\text{ext}}-\vb{f}^{0,\text{int}}}$\;
\While{$R>\epsilon$ and $n<n_{\text{max}}$}
{
\label{alg:dynamic_relaxation:loop_start}
Compute next time step\;
 $t^{n+1}\gets t^n+\Delta t^{n+1/2}$, $t^{n+1/2}\gets \frac{1}{2}\left(t^n+t^{n+1}\right)$ \;
Partial velocity update \tcp*{free dof vector operation}
$\vb{v}^{n+1/2} \gets \vb{v}^{n-1/2}+(t^{n+1/2}-t^{n}) \vb{a}^n$\;
Update displacements \tcp*{free dof vector operation}
 $\vb{u}^{n+1} \gets \vb{u}^{n}+\Delta t^{n+1/2} \vb{a}^n$\;
 Get internal forces from fiber constitutive equation\;
 $\vb{f}^{n+1,\text{int}}\gets$ \GIF{$\vb{u}$}\;
Compute damping force \tcp*{free dof vector operation}
$\vb{f}^{n+1,\text{damp}} \gets c\vb{m}\vb{v}$\;
Compute force residual  \tcp*{free dof vector reduction}
$R\gets \norm{\vb{f}^{n,\text{ext}}-\vb{f}^{n,\text{int}}}$\;
Update Accelerations \tcp*{free dof vector operation}
 $\vb{a^n+1} \gets 1/\vb{m}\left(\vb{f}^{n+1,\text{ext}}-\vb{f}^{n+1,\text{int}}-\vb{f}^{n+1,\text{damp}}\right)$\;
Partial velocity update \tcp*{free dof vector operation}
$\vb{v}^{n+1} \gets \vb{v}^{n+1/2}+(t^{n+1}-t^{n+1/2}) \vb{a}^{n+1}$\;
Optionally Update Energy and Check Energy Balance \tcp*{3 vector reductions}
Update iteration count\;
$n\gets n+1$
} \label{alg:dynamic_relaxation:loop_end}
\caption{Dynamic Relaxation Algorithm} 
\label{alg:dynamic_relaxation}
\end{algorithm2e}

Algorithm \ref{alg:dynamic_relaxation} outlines the basic dynamic relaxation algorithm used for the microscale RVEs. When used as part of the multiscale scheme, it computes macroscale stresses and material stiffness at every iteration of the macroscale solve (algorithm \ref{alg:macroFET}, line \ref{alg:macroFET:microsol}).

This dynamic relaxation algorithm is identical to a standard two-step central difference method, with the exception that the convergence criterion is based on a force residual measurement rather than time. Most of the operations in algorithm \ref{alg:dynamic_relaxation} are vector products. The entries of each vector are grouped by fixed and free degrees of freedom. This improves computational efficiency and cache utilization because most operations only need to operate on the free degrees of freedom. For example, once the displacements on the boundary nodes are set for a given deformation increment from the macroscale problem, their positions do not need to be recomputed for each iteration of the dynamic relaxation solver. They do however need to be stored, as their positions will contribute to the forces computed in \lstinline{getInternalForces} function. In algorithm \ref{alg:dynamic_relaxation} the group of degrees of freedom which each operation operates over is listed.

Algorithm \ref{alg:dynamic_relaxation} starts by loading the RVE mesh data and transforming the RVE connectivity graph into a flat array. Currently, the mass matrix is computed on the CPU, however, it only needs to be computed a single time for each RVE (over the course of a full multiscale analysis), so it does not represent a performance bottleneck at this time. In algorithm \ref{alg:dynamic_relaxation} superscripts denote the iteration. The algorithm proceeds by initializing the velocity, $\vb{v}$, displacements, $\vb{u}$, increment counter, $n$, and solution tolerance $\epsilon$. Next the internal and damping forces are computed and used to compute initial accelerations, $\vb{a}$, which are computed by multiplying the inverse of the diagonal stiffness matrix, $\vb{m}$ with the balance of external forces, $\vb{f}^{\text{ext}}$, internal forces, $\vb{f}^{\text{int}}$, and damping forces, $\vb{f}^{\text{damp}}$. A scalar damping constant $c$ is used to compute the damping forces. This corresponds to a mass weighted viscous damping of the nodes. Using the current multiscale methodology, the external forces are always zero, so they are not stored. The solution residual, $R$, is computed as the $L_2$ norm of the internal and external force imbalance. The solution proceeds by computing the accelerations, velocities, displacements, and forces in turn. Lines \ref{alg:dynamic_relaxation:loop_start}--\ref{alg:dynamic_relaxation:loop_end} iterate using the central difference method until the residual converges.

A two-step central difference method is used because it allows the system energy to be computed at full timesteps. Belytschko suggests that the system energy should be used to verify that an explicit simulation has not deviated from physical reality \cite{belytschkoNonlinearFiniteElements2014}. However, this is expensive because it requires 3 reductions of the whole system state during each iteration. Additionally, it is not clear that the energies help detect issues in the dynamic relaxation solve. This is because any invalid intermediate state is acceptable as long as the final damped state represents the static solution of the RVE boundary value problem. Other numerical conditioning issues are not guaranteed to be detected by the energy balance check. In effect, we seldom use the energy balance check.

An interesting aspect of the dynamic relaxation method is that since we are solving for the static state of the system, only the elasticity of the trusses is important to arrive at the correct solution; that is, the inertial parameters such as velocity, time, and mass are fictitious. That means, one can choose these parameters to minimize the time to solution (e.g., \cite{underwoodDynamicRelaxation1986,papadrakakisMethodAutomaticEvaluation1981,zhangDevelopmentMaDRMethod1994,rezaiee-pajandDynamicRelaxationMethod2010,leeSimpleExplicitArclength2011}). There are two main methods to accomplish this: kinetic damping and viscous damping. In this work, we make use of the viscous damping approach because it has a smaller memory footprint and memory throughput is one of the main bottlenecks with the GPU implementation. In our current implementation, we use a constant velocity damping coefficient and mass matrix because initial tests did not show improvement using an automatically updating damping coefficient, as often applied in the literature. However, this point needs further exploration because the appropriate damping coefficient for optimal performance changes as the RVE deforms and manually tuning the damping coefficient to work well in the analysis regime of interest is challenging.

One advantage of the dynamic relaxation algorithm is that since it is using blas-1 style vector operations, it permits a simple initial implementation on a GPU accelerator. To get adequate performance, a more careful analysis and implementation is required. The details are clarified in section \ref{sec:micro_software_arch}.

Another important aspect of our two-scale finite element method is the computation of the material stiffness. Since we use a dynamic relaxation method to solve the microscale problem, we do not have access to the Jacobian, or a guarantee that it is invertible at the deformation state where we seek to find the stiffness. To that end, we present the computation of the material stiffness making use of the finite difference directional derivatives.

The fourth order spatial tangent material stiffness tensor needed in an updated-Lagrangian scheme \(\vb{C}\) is not computed directly, but as the push forward of the derivative of the second Piola-Kirchhoff stress with respect to the Green-Lagrange strain \(A_{ijkl}=\pdv{\Pi_{ij}}{E_{kl}}\). Using Einstein summation notation, the push forward is written as
\begin{equation}
C_{mnpq} = \frac{1}{J} F_{mi} F_{nj} A_{ijrs} F_{pr} F_{qs},
\label{eq:stiffness-push-forward}
\end{equation}
where \(\vb{F}\) is the deformation gradient, and \(J=\det \vb{F} \) is the Jacobian.

So, the task of finding the updated-Lagrangian material stiffness becomes computing \(\vb{A}\), however the microscale solver takes the deformation gradient as an input, so we start by applying the chain rule as
\begin{equation}
\pdv{\Pi_{ij}}{E_{kl}} = \pdv{\Pi_{ij}}{F_{pq}}\pdv{F_{pq}}{E_{kl}} = \pdv{\Pi_{ij}}{U_{pq}}\pdv{U_{pq}}{E_{kl}}.
\end{equation}
In the third expression, we make use of the invariance of the second Piola-Kirchhoff stress to rotation. Critically, this invariance will limit the number of additional microscale solves needed to compute the derivatives to 6.

To compute \(\pdv{U_{pq}}{E_{kl}}\) we can make use of the inverse function theorem to write
\[\pdv{U_{pq}}{E_{kl}} = \left(\pdv{E_{pq}}{U_{kl}}\right)^{-1},\]
however this represents inversion of a fourth order tensor, and it is not initially obvious how to perform this inversion in practice. Fortunately, both \(\vb{E}\) and \(\vb{U}\) are symmetric, so the resulting fourth order tensor \(M\) will contain both minor symmetries. This permits the Mandel matrix representation of \(\vb{M}\) which conveniently preserves the inner product and can therefore be used to compute the tensor inverse directly.

In tensor form
\begin{align}
M_{klpq} = \pdv{E_{kl}}{U_{pq}} =\frac{1}{4} \left(
\delta_{kq}U_{pl} + \delta_{lq}U_{pk} +
\delta_{kp}U_{ql} + \delta_{lp}U_{qk}
\right).
\end{align}
This corresponds to the 6x6 Mandel matrix representation
\begin{equation}
   \begin{bmatrix}
       \vb{M}
   \end{bmatrix} = \frac{1}{2}\begin{bmatrix}
   2U_{11} & 0 & 0& 0 & \sqrt{2} U_{13} & \sqrt{2} U_{12} \\
   0 & 2U_{22} & 0 & \sqrt{2} U_{23} & 0 & \sqrt{2} U_{12} \\
   0 & 0 & 2U_{33} & \sqrt{2} U_{23} & \sqrt{2} U_{13} & 0 \\
   0 & \sqrt{2} U_{23} & \sqrt{2} U_{23} & U_{22} + U_{33} & U_{12} & U_{13} \\
   \sqrt{2}U_{13} &0 & \sqrt{2} U_{13} & U_{12} & U_{11}+U_{33} & U_{23} \\
   \sqrt{2} U_{12} & \sqrt{2} U_{12} & 0 & U_{13} & U_{23} & U_{11} + U_{22}
   \end{bmatrix}.
\end{equation}
In the remainder of this section, quantities in square braces are the 6x6 Mandel representation of 4th order tensor quantities.

The derivatives of the microscale stress are approximated using a first order directional derivative
\begin{equation}
    \pdv{\Pi_{ij}}{U_{rs}} T_{rspq} \approx  \frac{\Pi_{ij}(U_{lm}+h T_{lmpq})- \Pi_{ij}(U_{lm})}{h}=P_{ijpq},
    \label{eq:tensor_fd}
\end{equation}
where \(T_{rspq}\) is the direction tensor, \(h\) is a small perturbation and we have defined a 4th order tensor \(P_{ijpq}\) that holds the results of the finite difference operations in each direction. Note that \(p\) and \(q\) are free indexes that correspond to the probing direction and will be fixed in each evaluation of the finite difference. Due to the symmetry in the probing directions described below, this method only requires 6 finite difference evaluations. In practice, we only compute and store the 36 unique components of \(P_{ijpq}\).

Since, the stretch tensor \(U_{ij}\) is a symmetric positive definite matrix, care must be taken such that the directions are also symmetric positive definite. In this work, we chose the directions that are represented by the following matrix in the Mandel form:
\begin{equation}
    \begin{bmatrix}\vb{T}\end{bmatrix} = \frac{1}{2}\begin{bmatrix}
        1 & 0 & 0 & 0& \sqrt{2} & \sqrt{2} \\
        0 & 1 & 0 & \sqrt{2}& 0 & \sqrt{2} \\
        0 & 0 & 1 & \sqrt{2} & \sqrt{2} & 0 \\
        0 & 0& 0& 2& 0& 0 \\
        0 & 0& 0& 0& 2& 0 \\
        0 & 0& 0& 0& 0& 2
    \end{bmatrix},
\end{equation}
and emphasize that this choice is not unique. One can choose any set of directions that are both symmetric positive definite and span the 6 fundamental directions. The symmetric positive definite requirement prevents the use of a \textcolor{red}{orthogonal} set of directions.

Finally, we directly solve
\begin{equation}
    \left(\begin{bmatrix}\vb{M}\end{bmatrix}\begin{bmatrix}\vb{T}\end{bmatrix} \right)^T \begin{bmatrix}\vb{A}\end{bmatrix}^T = \begin{bmatrix}\vb{P} \end{bmatrix}^T
\end{equation}
for the tangent material stiffness \(A_{ijkl}\), and use equation \eqref{eq:stiffness-push-forward} to obtain the spatial tangent material stiffness.

\section{Software Architecture} \label{sec:software_arch}
\begin{figure}
    \centering
    \includegraphics[width=\linewidth]{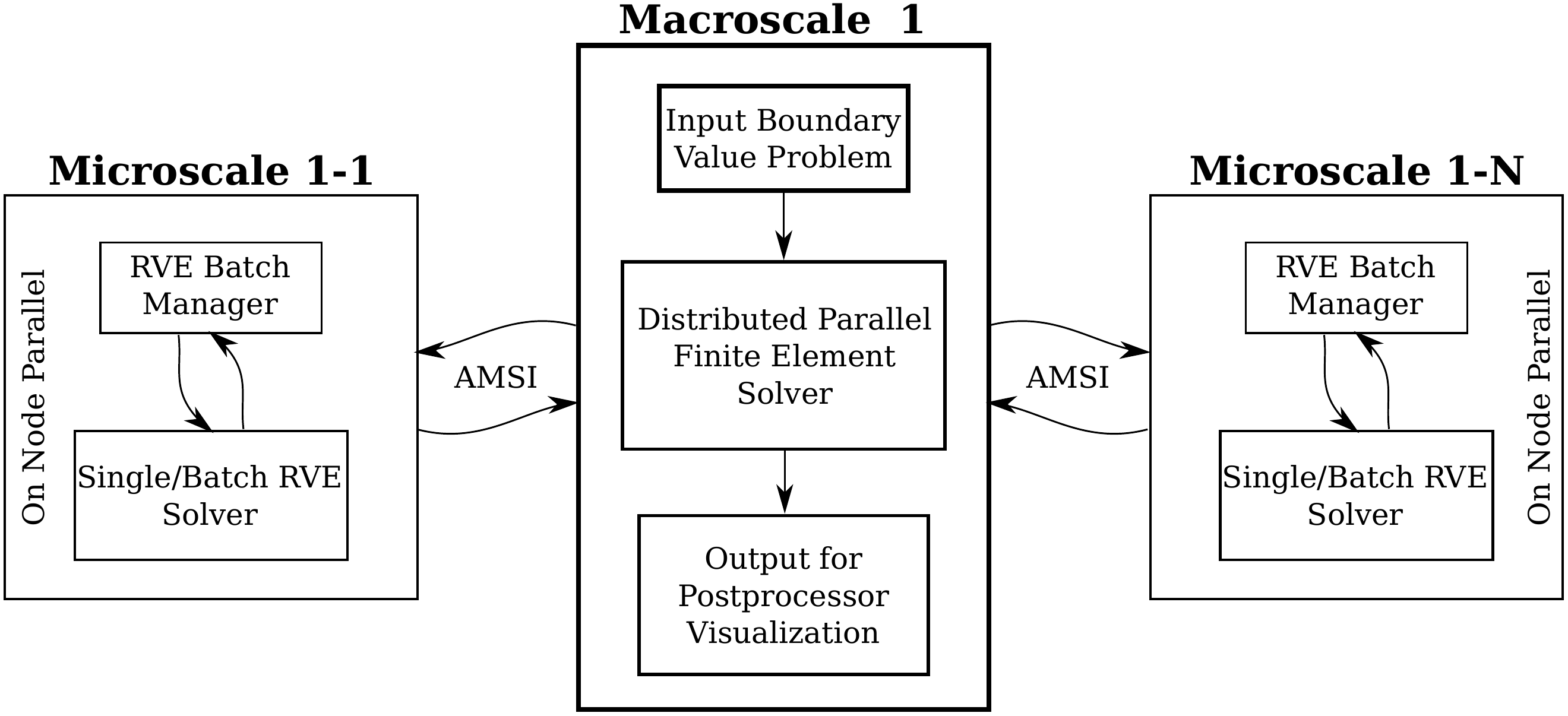}
    \caption{Structure of \MuMFiM{}.}
    \label{fig:multiscale_code_structure}
\end{figure}
Our analysis code is separated into three major components: \AMSI{}, \MuMFiM{}/Macroscale, \MuMFiM{}/Microscale. \AMSI{} is designed to be a general framework that enables the coupling of legacy single-scale analysis codes into a multiscale analysis. \MuMFiM{} is our main application which performs multiscale analysis of fibrous materials using the \FET{} method. Its macroscale component uses traditional FEM methods to solve for the static equilibrium of a Cauchy continuum. It is designed to be general and easy to add new constraints and materials. 

\textcolor{red}{To add a new macroscale material model, the user must only provide a function that computes the current stress and material stiffness based on the increment in strain. Like most modern FEM codes, this is done locally at each integration point. On the microscale, the procedure is similar, however, the microscale routines are ``batched'', that is they take a set of deformation increments and return the stresses and material stiffnesses for the entire set at once.
}

For the macroscale, it is assumed that there is a large mesh and many of its features revolve around this assumption. For the microscale, we assume that the mesh size of each RVE is small enough to fit on either a single processor or GPU. The combination of small RVE mesh sizes and RVEs comprised of truss networks offers significant optimization opportunities over the more general macroscale solver that supports a distributed mesh. Therefore, rather than using the same general finite element solver that we used for the macroscale on the microscale, we use specialized solvers that are optimized for the RVE of interest. New microscale solvers can be added as long as they abide by our batched analysis API.

\subsection{Scale Coupling}

The scale coupling design of \AMSI{} focuses on extensibility and ease of use. The goal of the design is to enable the use of previously developed single-scale analysis codes while developing a new multiscale analysis with minimal refactoring. We believe that many multiscale patterns can be implemented without forcing the re-development, re-validation, and re-verification of the single-scale physics. This service library approach is in contrast to the monolithic framework approach that is seen in many multiscale toolboxes such as MOOSE \cite{gastonMOOSEParallelComputational2009} and doesn't require adaption of internal data structures, or implementation of new APIs which is required in other multiscale toolboxes such as Uintah and MpCCI \cite{davisondestgermainUintahMassivelyParallel2000,joppichMpCCIToolSimulation2006}. The level of abstraction introduced by \AMSI{} does have a performance cost since unmodified application codes are not optimized for inter-scale coupling. In this sense, there is a trade-off between maximum parallel efficiency and the need for intrusive physics application modifications. The design of \MuMFiM{} has given a preference to minimal code modifications, trading computational efficiency for developer, verification, and validation efficiency.

\MuMFiM{} takes advantage of the multiple program multiple data (MPMD) paradigm by allowing multiple single-scale applications to be run simultaneously. To perform the communication between scales, we make use of MPI. Since multiple programs may be in flight simultaneously and most parallel applications expect to operate over the entire parallel execution space, we partition the parallel execution space into disjoint subsets. This choice is not ideal for certain types of coupling schemes, such as Gauss-Seidel, because it can leave a large portion of the compute resources idling while one scale waits for another to finish. This constraint will be relaxed in a later release of \MuMFiM{}. 

From the application programmer's perspective, these requirements are realized as a requirement that a program takes an MPI communicator as an input and does not perform operations over the global communicator \lstinline{MPI_COMM_WORLD}. The \AMSI{} library is responsible for partitioning the global parallel execution space and communicating between scales. Currently, there are two partitioning schemes available: \lstinline{Exclusive} and \lstinline{Strided}. The \lstinline{Exclusive} partitioning scheme gives $n_i$ contiguous processes to each of the $i$ scales. The \lstinline{Strided} partitioning scheme breaks up the process allocation over a given stride. Due to the challenges associated with the efficient use of global parallel operations, partitioning is the only global operation that is directly supported through the API. To give the developer freedom to directly use \lstinline{MPI} functionality, we provide the \lstinline{MUMFIM_COMM_WORLD} communicator which acts over the total global parallel execution space and \lstinline{MUMFIM_COMM_SCALE} which acts over the current scale's partition.

\MuMFiM{} provides two types of communication operations: reconciliations and assemblies. Reconciliations are inter-scale parallel operations. They are designed to be unidirectional and operate over the union of the two parallel execution spaces that are being reconciled. These operations are typically used to communicate simulation state between the analysis scales. One example of a reconcilliation is sending the new incremental deformation gradient to the microscale during each Newton step. Assemblies are intra-scale communications. The most common use case (and inspiration for the name) is assembling data on a scale before communicating.

To keep track of the scale coupling relationship, \AMSI{} makes use of a \lstinline{Coupling} data structure. The \lstinline{Coupling} data structure keeps track of which processes communicate with each other during an inter-scale communication. Each scale coupling represents a one-way communication. Although many multiscale analysis techniques require bi-directional communications, some do not \cite{gravemeierTaxonomyMultiscaleMethods2008}. Therefore, to achieve the most general interface, scale couplings are unidirectional. Bi-directional couplings can be achieved by composing two unidirectional couplings.

Listing \ref{lst:AMSITwoScale} shows how \AMSI{} can be used to set up a two scale analysis problem. The control service is a singleton that is used as a handle to set up the scale coupling. The \lstinline{setScaleMain} function on lines \ref{lst:AMSITwoScale:ln:setscalemain1}--\ref{lst:AMSITwoScale:ln:setscalemain2} takes a scale name and a function pointer and is used as the main entry point for that scale. The number of ranks on each scale and the couplings are set up in the \AMSI{} input file, which is read during the initialization function on line \ref{lst:AMSITwoScale:ln:init}. The relevant portion of an example \AMSI{} input for the code in listing \ref{lst:AMSITwoScale} is shown in listing \ref{lst:amsi_options}. In this example, the scale ``macro'' is set up with 10 ranks and the scale ``micro`` is set up with 99 ranks. The scale names in the input file must match the associated scale names in the analysis code. Here, a bi-directional coupling is established between the macro and micro scales in the \lstinline{@relations} section of the input file. With the default process allocator, the first 9 macroscale ranks will each communicate with 10 microscale ranks each and the 10th macroscale rank will communicate with 9 microscale ranks to make up the 99 total ranks.

\begin{lstlisting}[caption={Two-scale analysis setup with \AMSI{}.},label=lst:AMSITwoScale,language=c++,escapechar={|}]
using ScaleFuncPtr = int (*)(int&, char**&, MPI_Comm); |\label{lst:AMSITwoScale:ln:funcptr}|

int main_macro(int& argc, char** &argv, MPI_Comm comm) |\label{lst:AMSITwoScale:ln:macro_main}|
{
    // setup and run the macroscale analysis code
}
int main_micro(int& argc, char** &argv, MPI_Comm comm) |\label{lst:AMSITwoScale:ln:micro_main}|
{
    // setup and run the microscale analysis code
}

int main(int argc, char** argv)
{
    // Read amsi option file and set up the analysis.
    // This initializes MPI, and 
    // partitions the parallel execution space,
    // and defines the relation between each scale.
    initMultiscale(argc, argv, MPI_COMM_WORLD); |\label{lst:AMSITwoScale:ln:init}|
    // Get a control service handle. 
    auto control = ControlService::Instance();
    // Set the main function for each scale.
    // The setScaleMain function takes the scale name
    // and a function pointer with the type ScaleFuncPtr
    control->setScaleMain("macro", &main_macro); |\label{lst:AMSITwoScale:ln:setscalemain1}|
    control->setScaleMain("micro", &main_micro); |\label{lst:AMSITwoScale:ln:setscalemain2}|
    // Execute the analysis
    int result = control->Execute(argc, argv);
    // free MPI, perform any code cleanup
    freeMultiscale();
    
    return 0;
}
\end{lstlisting}
\begin{lstlisting}[caption={Scale setup in \AMSI{} scale coupling options file},label=lst:amsi_options]
@scales
macro 10
micro 99
@relations
macro micro
micro macro
\end{lstlisting}

Figure \ref{fig:multiscale_code_structure} shows the data flow for a single macroscale mesh partition. This general structure is repeated for each partition. A description of the macroscale inter-partition parallelization strategy is described below.

At the beginning of the multiscale analysis, the macroscale sets up a coupling topology based on the \AMSI{} input file. Each macroscale mesh partition resides on its own MPI rank. For the multiscale analysis, an RVE sub-scale problem must be solved at each integration point of the macroscale mesh. So, each macroscale rank communicates with up to $N$ microscale ranks, where $N$ can range from one to the number of integration points in the macroscale mesh partition. The optimal number of microscale ranks is dependent on the hardware topology (i.e., number and type of GPUs), and RVE size. A discussion of how many RVEs should be solved on each microscale rank is given in section \ref{sec:microscale_results}.

The macroscale analysis sends information about the RVE library that should be used at each integration point to the corresponding microscale analysis ranks. For many of the tissue modeling applications we have developed, the microscale rank randomly samples the RVE from the provided RVE library. However, other RVE assignment strategies such as explicitly assigning the RVE for each geometric region, or assigning from a background grid are available.

At each increment in the macroscale solver, the coupling can be updated to migrate the microscale RVEs between microscale ranks. This is helpful to account for migration of macroscale elements across macroscale ranks during mesh adaption and forl oad balancing between the microscale ranks within each macroscale mesh partition. Currently, this functionality isn't available for the batched microscale analysis which are used with the GPU accelerated version of the microscale code. This microscale migration is expected to be added in future versions of the batched microscale code.

After the coupling is updated and RVEs have been migrated, the macroscale analysis sends the increment in deformation gradient to the microscale. After the microscale solver computes the stress and material stiffness, it sends that data back to the relevant macroscale rank. Additional data such as the orientation tensor of the microscale RVE is sent to the macroscale to write into the output database on each converged macroscale step.

\subsection{Macroscale Structure and Parallelization} \label{sec:macro_software_arch}
The macroscale portion of \MuMFiM{} uses a modular object-oriented design for constructing the updated-Lagrangian finite element solver. The solver makes use of \PUMI{}, which provides the meshing and field database \cite{ibanezPUMIParallelUnstructured2016}. The solver interacts with two parallel components: a distributed parallel finite element solver and coupling data to the microscale solve through \AMSI{} as originally described in \cite{tobinAdaptiveMultiscaleSimulation2018}. When the macroscale solver is used as a standalone single-scale mode application, parallel coupling is not needed.

The macroscale solver builds upon the distributed meshing infrastructure provided by \PUMI{} \cite{ibanezPUMIParallelUnstructured2016}. One of the advantages of writing a FEM application on top \PUMI{} is that the mesh data structures maintain a link to the geometric model definition \cite{ibanezPUMIParallelUnstructured2016}. Since there was a lack of open source libraries for model setup which maintain the geometric relationships, an open source library, \textsc{model-traits}, was developed in parallel with \MuMFiM{} which provides an API to set up and apply boundary conditions based on the model geometry \cite{mersonModeltraitsModelAttribute2021}. This boundary-condition library can interface with Simmetrix SimModeler so the user can set up \MuMFiM{} models in a GUI environment.

\PUMI{} also provides tools for mesh partitioning, mesh adaption, and parallel mesh, geometry, and field interrogation. Within this framework, the macroscale mesh is partitioned into ``parts'' with a single part residing on each MPI rank. Solution field values on the boundaries of each part are synchronized after each iteration of the macroscale solver. \PUMI{} internally manages the (MPI) communication for field values on the part boundaries and \MuMFiM{} manages the synchronization through calls to the \PUMI{} API.

\MuMFiM{} uses \PUMI{}'s field integration framework to perform the element-wise tangent-stiffness matrix and force calculations. The elemental tangent-stiffness matrices and forces are assembled into a global PETSc sparse parallel matrix and vector, respectively. These parallel linear algebra structures are partitioned in the manner as the mesh; that is, the portions of these structures residing in a given rank refer to the same degrees of freedom that the mesh owns on that rank. This design means that new materials can easily be added by writing the element integration routines for the new material type through a \PUMI{} \lstinline{Integrator} class.

\MuMFiM{} provides a convenient abstraction for composing numerical solvers through inheritance. This abstraction is based on the idea that most solvers consist of a loop structure where some task is completed and some convergence criterion is computed on each loop iteration. To this end, the numerical solve function takes a pointer to an iteration object and a convergence object which provide methods to compute the per iteration task, and convergence criteria respectively.

This dynamic polymorphism works well for the macroscopic scale where a significant amount of work such as a linear solve is performed on each iteration, dwarfing the cost of the virtual function calls. However, the function call overhead is too large to use the same technique with the dynamic relaxation method that is used in the microscale solver. Due to this overhead, the solver loop is currently not abstracted in the microscale solver. Future versions of \MuMFiM{} expect to alleviate this issue by making use of a templated form of the \lstinline{numericalSolve} function which takes functor objects.

\subsection{Microscale Structure and Parallelization} \label{sec:micro_software_arch}
The \MuMFiM{}/Microscale is broken into two main pieces. The first is the driver component referred to as the RVE batch manager, which resides on a CPU and is responsible for communication, data marshalling, selecting the appropriate solver to run, and other tasks which are not GPU accelerated. The second is the solver component, which computes the stress and material stiffness for a given increment in the deformation gradient. Figure \ref{fig:multiscale_code_structure} gives a schematic representation of these components.

The majority of the floating point operations (FLOPS) for the multiscale analysis are associated with the microscale RVE calculations. Since more than 95\% of the pre-exascale and exascale supercomputers' available FLOPS are obtained from the GPU accelerators, we chose to port the microscale RVE calculations to GPUs. We have made use of the Kokkos performance portability layer \cite{carteredwardsKokkosEnablingManycore2014,trottKokkosProgrammingModel2022}.

In the naive approach, the dynamic relaxation algorithm is carried out using fused kernels for any subsequent operations with the same loop characteristics. The benefits of kernel fusion have been discussed extensively in the literature both for the case of explicit ODEs, and general GPU computations \cite{korchAcceleratingExplicitODE2018,wangKernelFusionEffective2010,wahibScalableKernelFusion2014}. The \lstinline{getInternalForces} subroutine accounts for two Kernel launches: the first to zero the internal force vector, and the second to scatter the elemental internal forces to the nodes. The current implementation uses atomic operations to scatter the forces.

In this naive approach, a number of microscale RVEs were assigned to each MPI rank and were executed serially with respect to each other within each rank. Despite the use of GPU acceleration for the vector operations, this approach had speedups less than one for the RVEs with less than ~3000 degrees of freedom when compared with a CPU-only implementation with serial vector operations. To unintrusively improve this naive approach, NVIDIA MPS was used to allow kernels from multiple MPI ranks to run concurrently. The use of MPS led to significant performance improvements for small DOF problems compared with the naive case. When using MPS, the problem size and number of simultaneous MPI ranks that are used can have a drastic effect on performance. All MPS results presented in section \ref{sec:microscale_results} use 32 MPI ranks per GPU, which gives the best performance in the range of problem sizes discussed here.

Since the loop in algorithm \ref{alg:dynamic_relaxation} executes millions of times per macroscale simulation step, we observed that this approach had significant kernel launch overhead. To overcome this, we moved the entire loop into a single kernel. This was done using Kokkos hierarchical parallelism, which uses teams of threads to enable a 2D map to the hardware. The CUDA reciprocal to this mechanism is launching a 1D grid of 1D blocks. Since our sub-scale problems each have less than 10,000 free degrees of freedom, we found that good performance could be achieved by assigning one thread team to each sub-scale problem. Here, we juxtapose the free degrees of freedom which are those without any Dirichlet constraints, to what we call degrees of freedom which are all potential degrees of freedom. Unlike an implicit FEM method, the constrained degrees of freedom cannot be eliminated as they are needed for the internal force computation. Reordering the fixed degrees of freedom to a contiguous block at the end of the displacement array allows most of the update algorithm to only operate on the smaller proportion of free degrees of freedom (algorithm \ref{alg:dynamic_relaxation}).

The choice of number of threads per team had a strong effect on performance. The ideal number of threads per team is a function of the microscale problem size. All presented results use 512 threads per team, which provided a good compromise for the performance of the smallest and largest RVEs we tested. 

A \lstinline{PackedView} data structure which has similar semantics to a Kokkos provided \lstinline{DualView} was used to allow effective access to multidimensional vector data within each thread team \cite{mersonKokkospackeddata2020}. This data structure uses a row vector and value vector, similar to those from compressed row storage (CRS), to store the data associated with all sub-scale problems on the current MPI rank in a contiguous array in memory. Each sub-scale problem gains access to the correct portion of memory through a Kokkos \lstinline{Subview}. In some ways, this structure is similar to a Kokkos \lstinline{View} of \lstinline{Views}. However, with the current implementation, the \lstinline{PackedView} can not be resized after initialization. A comprehensive performance comparison between the \lstinline{PackedView} data structure, and \lstinline{View} of \lstinline{Views} has not been performed to date. This differs from the \lstinline{StaticCrsGraph} in Kokkos which cannot handle non-integral datatypes and does not have \lstinline{DualView} semantics.

Although moving the analysis loop inside a single kernel launch was effective for our problems of interest, it can easily succumb to low performance from high register pressure. Significant effort had to be made to reduce the register pressure and ensure that multiple warps could be concurrently scheduled. One mechanism we used to reduce register pressure was to move some of the variables which are carried across loop iterations, such as the pseudo-time and the loop iteration count, into shared memory. We found that performance gain from the reduction in register pressure outweighed the loss in bandwidth from moving these variables to shared memory. The need to reduce register usage in this single kernel implementation led us to favor a stripped-down version of our algorithm which was specific to the physical system at hand. In other words, the flexibility of our code had to be sacrificed to obtain improved performance characteristics.

\section{Performance Results} \label{sec:PerformanceResults}
\subsection{Standalone Microscale Performance} \label{sec:microscale_results}

The performance results, presented in this section, are all computed on a single Volta V100 GPU---part of an IBM AC922 node. The code is compiled with version 16.1.0 of IBM's XL compiler for host code, version 10.1 of CUDA, and version 3.1 of Kokkos. The MPS results make use of Spectrum MPI version 10.3.

\begin{figure}
    \centering
    \includegraphics[width=0.5\linewidth]{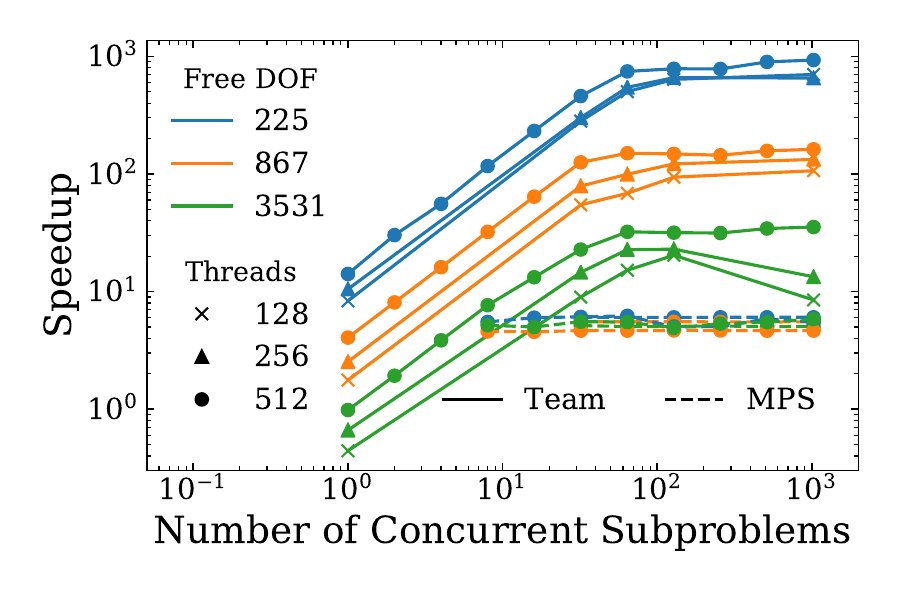}
    \caption{Speedup of the team-based (solid line) and MPS-based (dashed line) analysis over the naive approach. The MPS lines correspond to launching GPU kernels from 32 MPI ranks simultaneously. Each data point is the mean of three analysis runs.}
    \label{fig:speedup}
\end{figure}

Figure \ref{fig:speedup} shows the speedup of the thread team-based and MPS based analysis methods over the naive loop-based approach. In this plot, we see an initial linear scaling and a plateau region. We observe that as the problem size increases, the speedup obtained from the team thread-based method decreases. This is likely due to a reduction in the percentage of the problem which resides in the cache. We also observe that MPS based parallelism does provide some speedup over the naive approach, but it is not as effective as the thread team-based approach. Also, for MPS, the speedup plateau does not depend on the problem size. One way to interpret these results is that for maximum efficiency, at least 80 concurrent sub-problems should be run on each GPU. Since the V100 has 80 SMs (streaming multiprocessors), this is consistent with each Kokkos team (CUDA block) occupying a single SM.

The speedups achieved using thread team parallelism tell a compelling story that moving an analysis loop inside a single heavyweight kernel can be an effective optimization mechanism for problems that need to solve many problems, which cannot saturate the GPU on their own. Although MPS seemed like it might be a reasonable solution, it suffered from incurring a high kernel call latency due to the many kernels which were being called inside a hot loop. Additionally, the MPS solution was not able to make as effective use of the cache since many sub-scale problems were competing to be scheduled simultaneously, and each sub-scale problem that was scheduled in an interleaved fashion would cause cache misses.

\begin{figure}
\centering
\begin{subfigure}[t]{0.48\textwidth}
    \centering
    \includegraphics[width=\linewidth]{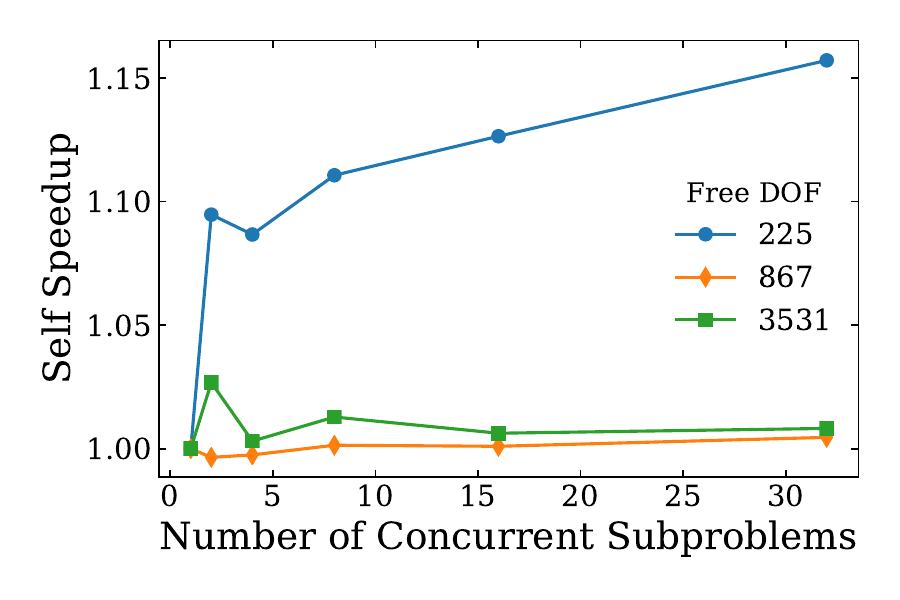}
    \caption{Speedup of the naive loop based analysis normalized by the number of concurrent sub-problems compared with the loop based single sub-problem. A flat line corresponds to a linear increase in runtime. Each data point is the mean of three analysis runs.}
    \label{fig:loop_self_speedup}
\end{subfigure}
\hfill
\begin{subfigure}[t]{0.48\textwidth}
    \centering
    \includegraphics[width=\linewidth]{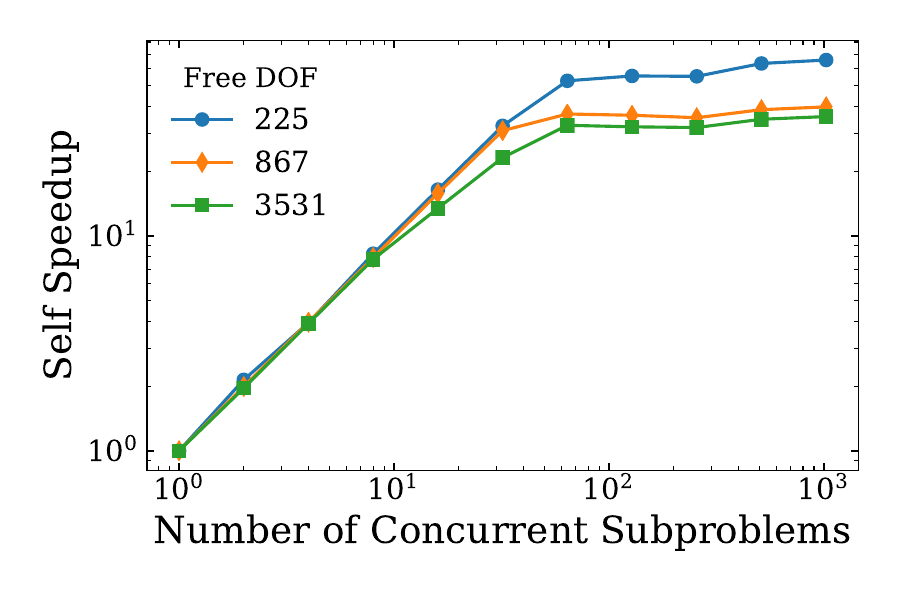}
    \caption{Speedup of the thread team based analysis normalized by the number of concurrent sub-problems compared with the thread team based single sub-problem. Each data point is the mean of three analysis runs.}
    \label{fig:team_self_speedup}
\end{subfigure}
\caption{Self speedup of the \MuMFiM{} microscale solver.} \label{fig:self-speedup}
\end{figure}

Figure \ref{fig:self-speedup} shows the runtime of a single sub-problem divided by runtime normalized by the number of concurrent sub-problems. This gives a measure of the speedup of a single sub-problem when computed in a concurrent batch. Since we are using the analysis technique's own single sub-problem runtime as a baseline for the speedup, we call this the ``self speedup''. For the naive loop based case (figure \ref{fig:loop_self_speedup}), we see the self speedup is flat which indicates the expected linear increase in runtime. The smallest problem size sees a slight self speedup. When thread team based parallelization is used, a significant self speedup is observed (figure \ref{fig:team_self_speedup}). Here we see an initial regime of linear self speedup and a plateau regime for large numbers of concurrent sub-problems. In this initial linear scaling regime, the runtime remains flat since the numerical workload is not large enough to overcome the kernel launch latency. Interestingly, the initial self speedup is almost identical for each of the problem sizes we tried. The plateau region shows that as the number of degrees of freedom in the problem increase, the self speedup decreases.

\subsection{Overall Parallel Performance} \label{sec:mumfim_performance_results}
The performance results in this section are computed on the AiMOS supercomputer at RPI. It has 252 IBM AC922 nodes each containing 2, 20 core IBM power 9 processors clocked at 3.15GHz, 512 GiB of RAM, and 6 Volta V100 GPUs. \MuMFiM{} was compiled with gcc 8.4.1, CUDA 11.1, Kokkos 3.2.0, and Spectrum MPI 10.4. For the macroscale solver, \MuMFiM{} makes use of PETSc 3.13.0, Hypre 2.18.2 (for the iterative solver), and \PUMI{} 2.2.5. 

\begin{figure}
    \centering
    \begin{subfigure}[b]{0.48\textwidth}
    \centering
        \includegraphics[width=\textwidth]{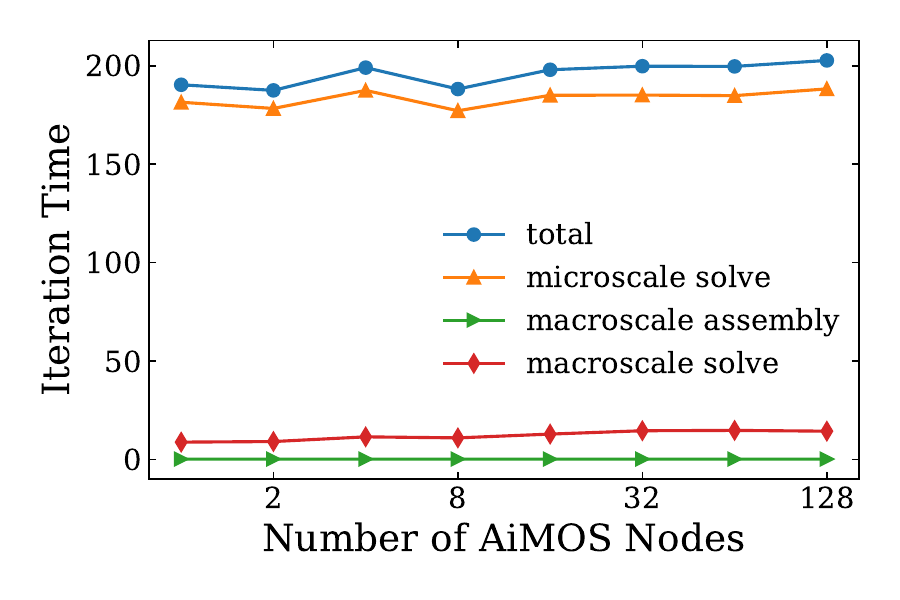}
        \caption{Weak scaling of \MuMFiM{}.}
        \label{fig:mumfim_weakscale}
    \end{subfigure}
    \hfill
    \begin{subfigure}[b]{0.48\textwidth}
        \centering
        \includegraphics[width=\textwidth]{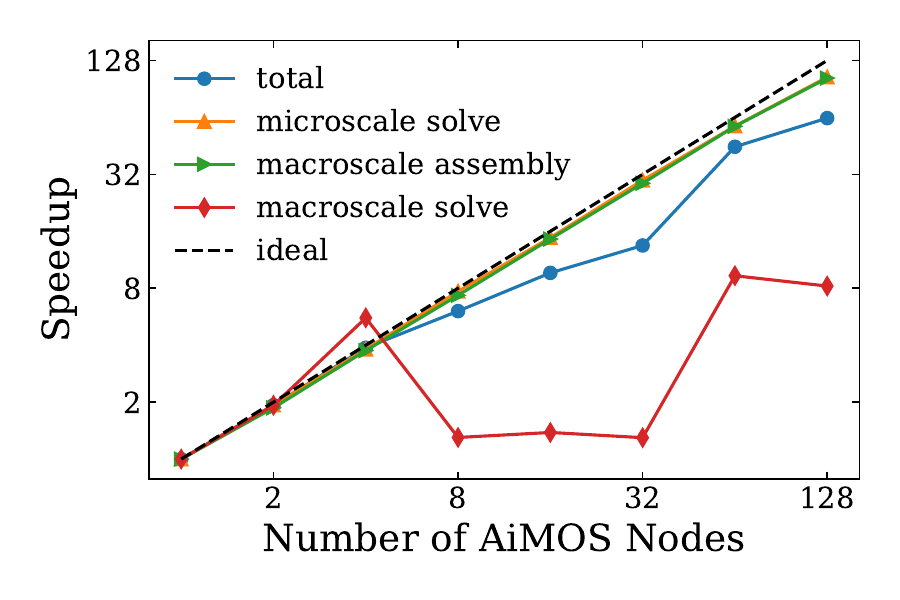}
        \caption{Strong scaling of \MuMFiM{}.}
        \label{fig:mumfim_strongscale}
    \end{subfigure}
    \caption{Scaling results for \MuMFiM{}.}
\end{figure}

Figure \ref{fig:mumfim_weakscale} shows the weak scaling of \MuMFiM{} for various components of the solution procedure. For this result, one macroscale rank and six microscale ranks, one for each GPU, is allocated on each node. The iteration runtime is taken from the second macroscale iteration to remove the effect of initialization. The mesh is constructed from linear tetrahedrons and contains 6080 elements, approximately 1000 RVEs per GPU. As the number of AiMOS nodes is doubled, the number of elements in the mesh is also approximately doubled maintaining a ratio of approximately 6000 microscale RVEs per macroscale rank. The macroscale linear solve uses the parallel incomplete lu solver (PILUT) as a preconditioner to the iterative conjugate gradient solver.

The weak scaling results in figure \ref{fig:mumfim_weakscale} show ideal weak scaling for each of the components of the multiscale procedure. The solution of the microscale RVEs marked with upward pointing triangles took an order of magnitude more time than the macroscale solve (marked with diamonds). This result indicates that future optimizations should focus on reducing the microscale solution time over further improving the macroscale element assembly or macroscale parallel linear solver.

Strong scaling of \MuMFiM{} was performed using a mesh with 189,140 linear tetrahedron elements. These results also use a single macroscale rank and six microscale ranks on each node. Figure \ref{fig:mumfim_strongscale} shows the strong scalability as a speedup plot. The solution of microscale RVEs exhibits nearly ideal strong scaling. This result is expected because each RVE can be computed independently. The macroscale assembly process also exhibits ideal strong scaling. The macroscale solver has very poor strong scaling. This isn't surprising since the macroscale portion of the problem is quite small with only 105,861 unconstrained degrees of freedom. These results indicate that for this problem size it is not worthwhile to use more than 4 ranks for the macroscale solve. Although the macroscale FEM assembly exhibits strong scalability beyond 4 ranks, it encompasses such a small percentage of the total solve time that reducing it makes effectively no improvement in the time-to-solution.

\section{Example Application: Modeling of the Facet Capsular Ligament} \label{sec:ExFCL}

\begin{figure}
    \centering
    \begin{subfigure}[t]{0.48\textwidth}
        \centering
        \includegraphics[width=\textwidth]{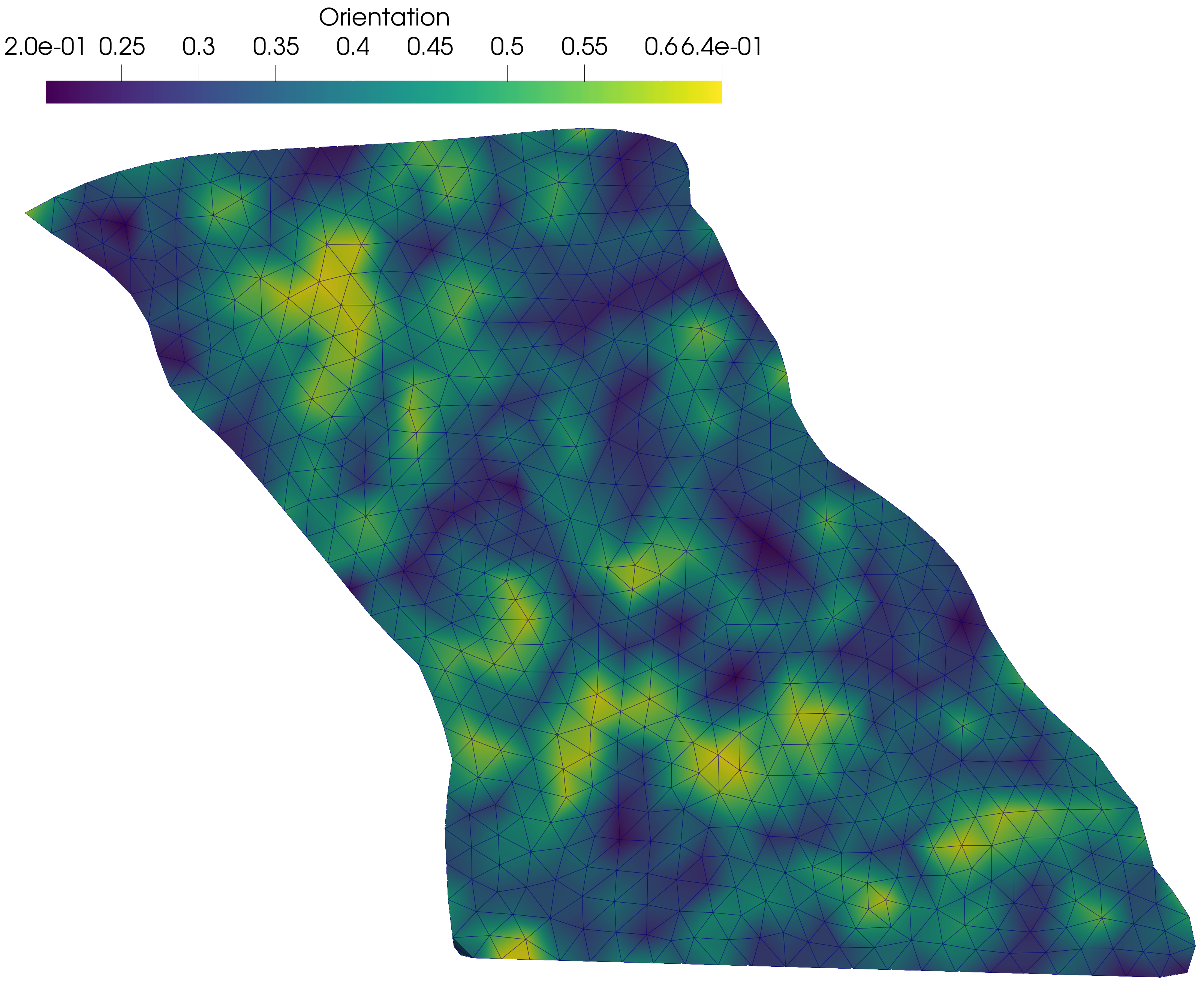}
        \caption{Initial orientation of microscale collagen fibers.}
    \label{fig:fcl_model_setup}
    \end{subfigure}
    \hfill
    \begin{subfigure}[t]{0.48\textwidth}
        \centering
        \includegraphics[width=\textwidth]{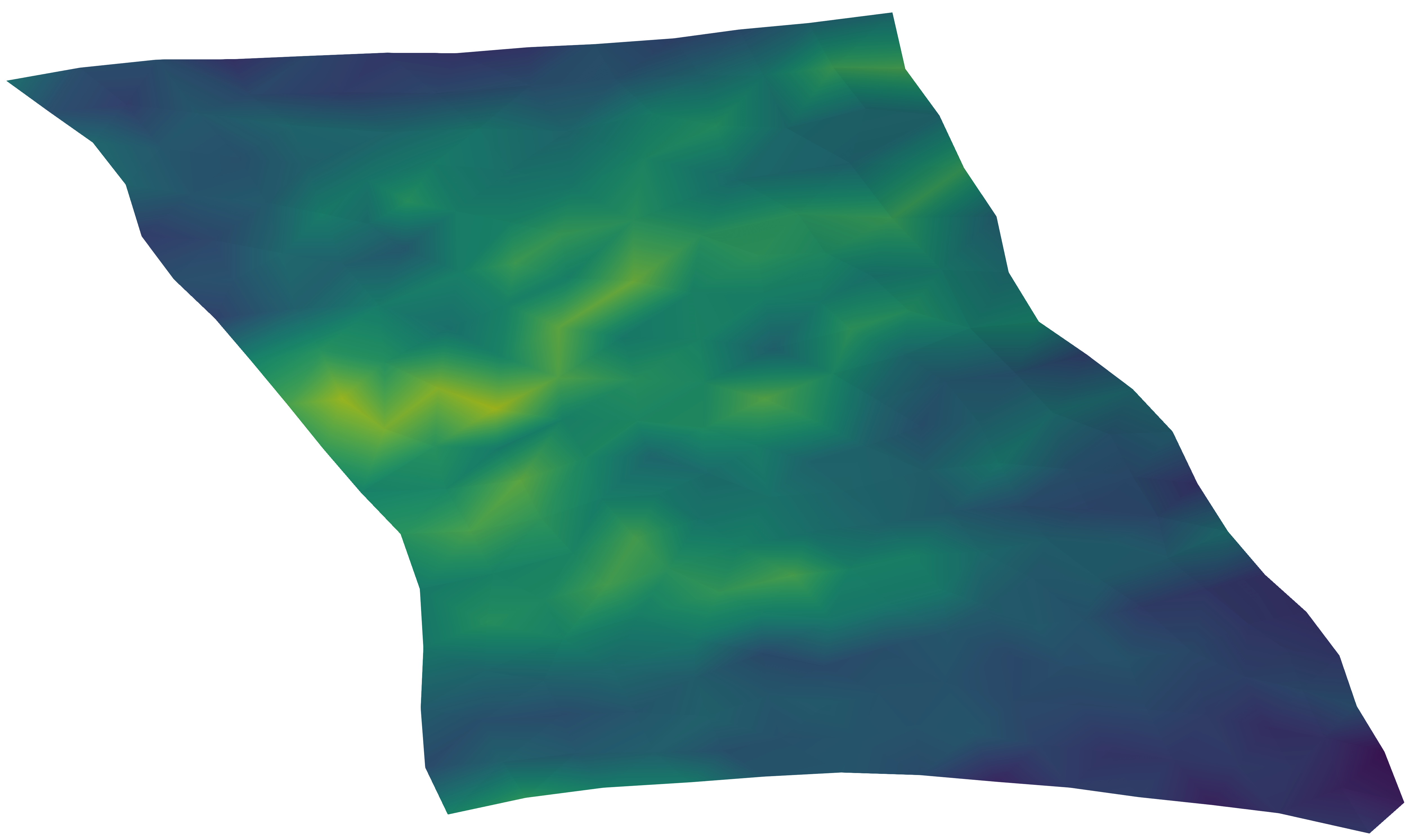}
        \caption{Representative stress field after 50\% strain.}
    \label{fig:fcl_stress_field}
    \end{subfigure}
    \caption{Facet capsule ligament (FCL) model with approximately 26 thousand linear tetrahedron elements.}
\end{figure}

The facet capsular ligament (FCL) is a ligament that connects the spinal processes. Understanding the relation between the fibrous collagen structure and the mechanical behavior of the ligament is an area of active research in the biomedical engineering community \cite{corralesImportanceCervicalCapsular2021,bermelAsymmetricInplaneShear2020,gacekThroughthicknessRegionalVariation2021,itaIntraarticularCollagenaseSpinal2020}. This research is critical because the FCL is believed to be an important contributor to back and neck pain, which are experienced by 30\%-50\% of the population during their lifetime and contributes to a yearly cost of \$30 billion dollars \cite{singhPhysiologicFacetCapsule2019,cohenEpidemiologyDiagnosisTreatment2015,coteAnnualIncidenceCourse2004,hoyEpidemiologyNeckPain2010}.

The FCL is primarily made up of type I collagen \cite{banCollagenOrganizationFacet2017}. Experimental works have shown that reorientation of the collagen ECM  during loading is responsible for causing pain in the FCL \cite{zareiTissueLoadingMicrostructure2017}. This fiber level reorientation is not possible to capture using homogenized single-scale continuum models such as the Holzapfel-Gasser-Ogden (HGO) model. Additionally, collagen fiber networks exhibit significant nonaffine behavior which cannot be modeled with homogeneous Cauchy-Continuum material models \cite{hatami-marbiniScalingNonaffineDeformation2008,huismanInternalStressesNormal2011,picuMechanicsRandomFiber2011,chandranAffineNonaffineFibril2006,mersonSizeEffectsRandom2020}. Lastly, the type I collagen fibers are about 500 nm in diameter, which is roughly five orders of magnitude smaller than the scale of the biological structures of interest \cite{shermanMaterialsScienceCollagen2015}.

Previous work done on modeling of fibrous materials has primarily used either discrete network models, or developed constitutive descriptions by fitting of continuum models to experimental data \cite{grekasCellsExploitPhase2019,hatami-marbiniMultiscaleModelingSemiflexible2013, islamStochasticContinuumModel2018,picuMechanicsRandomFiber2011,picuPoissonContractionFiber2018,deogekarStrengthRandomFiber2018,shahsavariSizeEffectMechanical2013,licupElasticRegimesSubisostatic2016,licupStressControlsMechanics2015,headMechanicalResponseSemiflexible2005,levineDeformationFieldSemiflexible2004}. Barocas et al. asserts a unique multiscale methodology which includes a modified macroscale equilibrium equation with an additional body force due to microscale fluctuations \cite{chandranDeterministicMaterialBasedAveraging2007,stylianopoulosVolumeaveragingTheoryStudy2007}. However, careful evaluation of the additional term shows that it must be equal to zero, which makes their method functionally equivalent to \FET{}.
Barocas et al. has made use of their method to study fibrous biological structures such as the arterial wall, cell contraction in a collagen gel, and the FCL \cite{aghvamiMultiscaleMechanicalSimulations2013,stylianopoulosMultiscaleStructurebasedModeling2007,zareiImagebasedMultiscaleMechanical2017}. 
However, these previous works have been limited to small mesh sizes at the macroscale, and small RVE sizes at the microscale.

Although \MuMFiM{} makes use of some of the finite element infrastructure developed in \cite{chanImagebasedMultiscaleMechanical2019,tobinAdaptiveMultiscaleSimulation2018} it has emerged as a new project.
Some of the most significant improvements in \MuMFiM{} include use of a new mathematical coupling formulation, new data-parallel algorithms for the microscale solution, improved geometric input file handling \cite{mersonModeltraitsModelAttribute2021} and more accurate microscale material stiffness calculations.
Together, these advancements have allowed for macroscale and microscale models which are two orders of magnitude larger than previously possible, and the solution of global strains of more than 50\%.

Ban et. al. observed that the cervical FCL contains regions of similarly oriented collagen fibers \cite{banCollagenOrganizationFacet2017}. It is suspected that these regions of oriented fibers may contribute to important mechanical processes such as the onset of collagen fiber reorientation, which has been experimentally linked to pain \cite{zareiTissueLoadingMicrostructure2017}. A complete description of our hypotheses and the modeling work required to explicate them is beyond the scope of this paper. Instead, we will provide examples of the types of information we can collect and a representative problem setup and forgo the detailed discussion of their biological relevance.

The front view FCL geometry is shown in figure \ref{fig:fcl_model_setup}. It is colored by the local orientation of the fibers in the initial undeformed configuration. This problem setup is designed to mimic the patches of aligned fibers observed in the cervical FCL reported in \cite{banCollagenOrganizationFacet2017}. This geometry is meshed with approximately 26,000 linear tetrahedron elements. The left-hand face of the FCL geometry has fixed displacements in the horizontal direction, and otherwise has traction free boundary conditions. A single point on the left-most face is held encastre to prevent rigid body modes. The right-most face of the FCL has displacement applied in the horizontal direction until the strain reaches roughly 25\%.

The normal stress in the horizontal direction is shown in figure \ref{fig:fcl_stress_field} for a model undergoing 50\% strain. Here we see localization bands forming across the ligament. These localization bands do not form in homogeneous elastic materials but are characteristic of fibrous materials. The regions of intense stress are more likely to fail and may cause neuron activation and pain.
\begin{figure}
    \centering
    \begin{subfigure}[t]{0.32\textwidth}
        \centering
        \includegraphics[width=\textwidth]{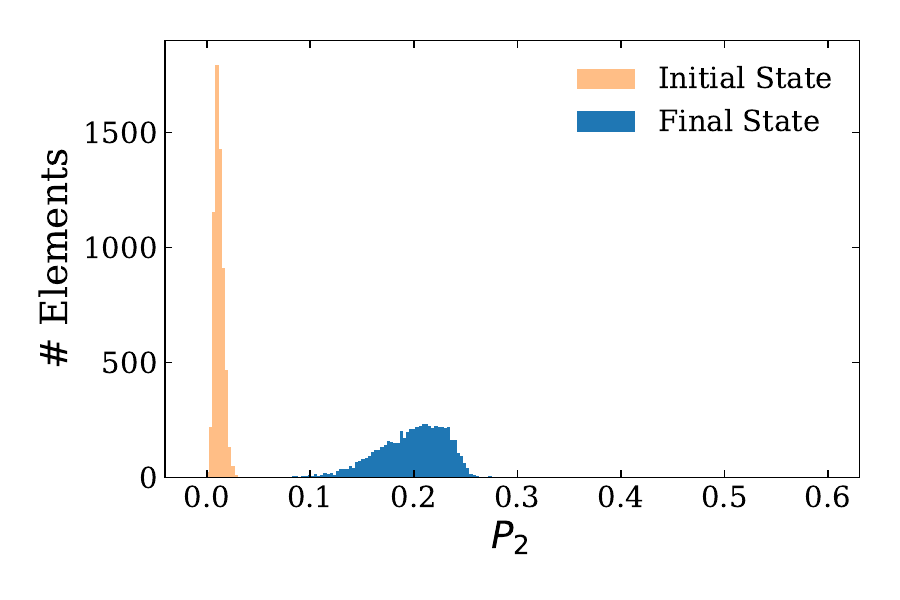}
        \caption{Case 1: Unaligned network.}
        \label{fig:degree_of_alignment:unaligned}
    \end{subfigure}
    \hfill
    \begin{subfigure}[t]{0.32\textwidth}
        \centering
        \includegraphics[width=\textwidth]{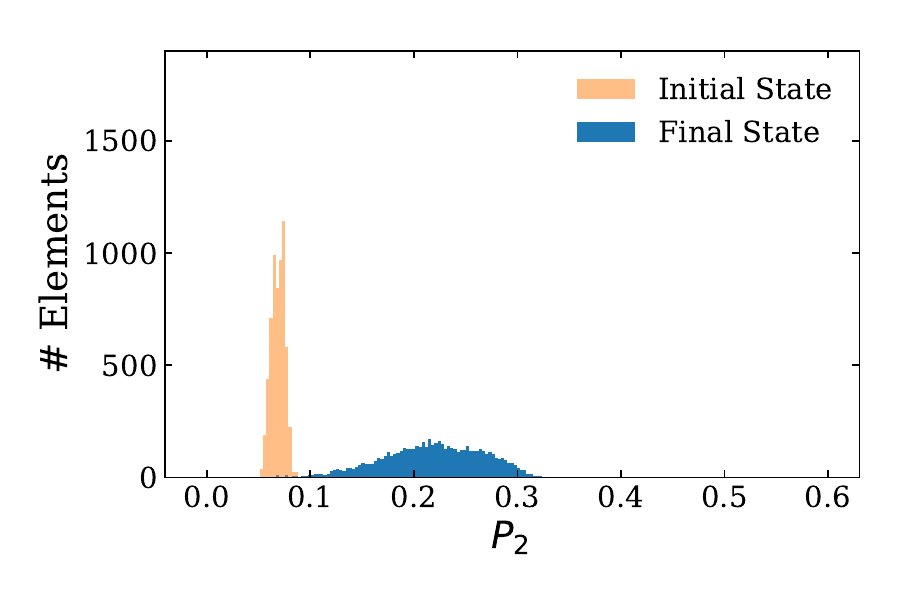}
        \caption{Case 2: Moderate alignment.}
        \label{fig:degree_of_alignment:moderate}
    \end{subfigure}
    \hfill
    \begin{subfigure}[t]{0.32\textwidth}
        \centering
        \includegraphics[width=\textwidth]{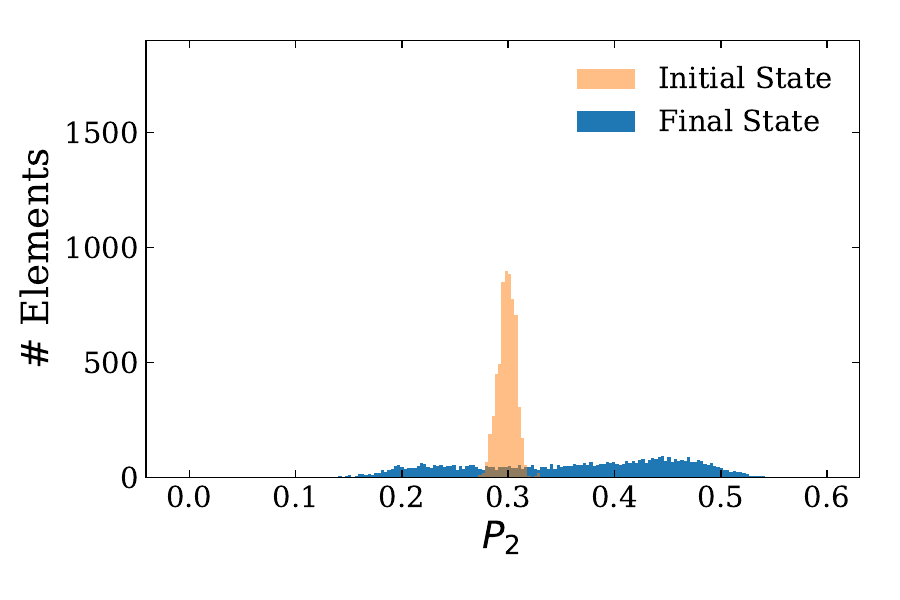}
        \caption{Case 3: Highly aligned.}
        \label{fig:degree_of_alignment:high}
    \end{subfigure}
    \caption{Initial and final degree of alignment for RVEs with patches of oriented networks.}
    \label{fig:degree_of_alignment}
\end{figure}

One of the benefits of \MuMFiM{} is that it can track the reorientation process of the collagen fibers and, in general, the evolution of the microstructure during loading. Figure \ref{fig:degree_of_alignment} shows how the degree of alignment shifts during the uniaxial extension process. The degree of alignment is measured by the second Legendre polynomial,
\begin{equation}
    P_2=\frac{1}{2}\langle 3 \cos^2{\theta} -1 \rangle,
\end{equation}
where $\theta$ is the angle between the fiber direction vector and a reference direction, which is typically taken as the loading direction. Angular parentheses represent averaging over the fibers in the RVE. This measure ranges between $-1/2$ and $1$. It is $-1/2$ when the fibers are aligned perpendicular to reference direction and $1$ when fibers are aligned in reference direction. A value of 0 represents a network where the direction vectors of fibers are uniformly distributed around the unit sphere.

Three states of initial alignment are shown in orange in figure \ref{fig:degree_of_alignment}. In each case, the initial alignment is a peak centered around the desired $P_2$ value. In the moderate and highly aligned cases (figures \ref{fig:degree_of_alignment:moderate} and \ref{fig:degree_of_alignment:high}), the orientation of the RVEs follows the stochastic orientation field shown in figure \ref{fig:fcl_model_setup}. In each case, the distribution of RVE alignments is much broader in the final state compared with that of the initial state. Also, a significant number of RVEs orient in the loading direction. Interestingly, in the case with the highest initial alignment, figure \ref{fig:degree_of_alignment:high}, some RVEs orient away from the loading direction creating a very wide distribution of RVE orientations. We are currently using \MuMFiM{} to understand the role that these RVE fiber re-orientations play in the biomechanics of the FCL.
\begin{figure}
    \centering
    \begin{subfigure}[t]{0.48\textwidth}
        \centering
        \includegraphics[width=\textwidth]{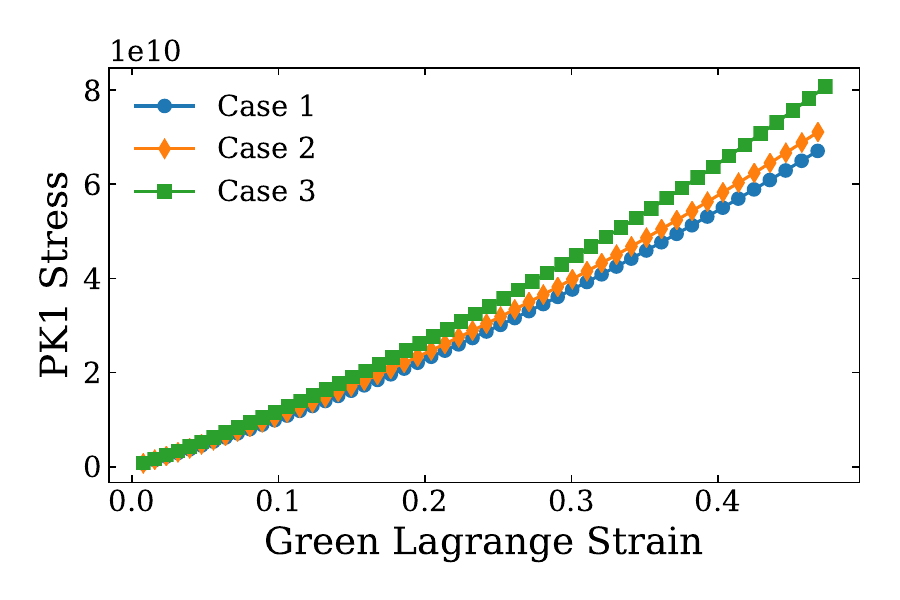}
        \caption{Stress strain curve of FCL with various initial orientation fields.}
        \label{fig:stress_strain}
    \end{subfigure}
    \hfill
    \begin{subfigure}[t]{0.48\textwidth}
        \centering
        \includegraphics[width=\textwidth]{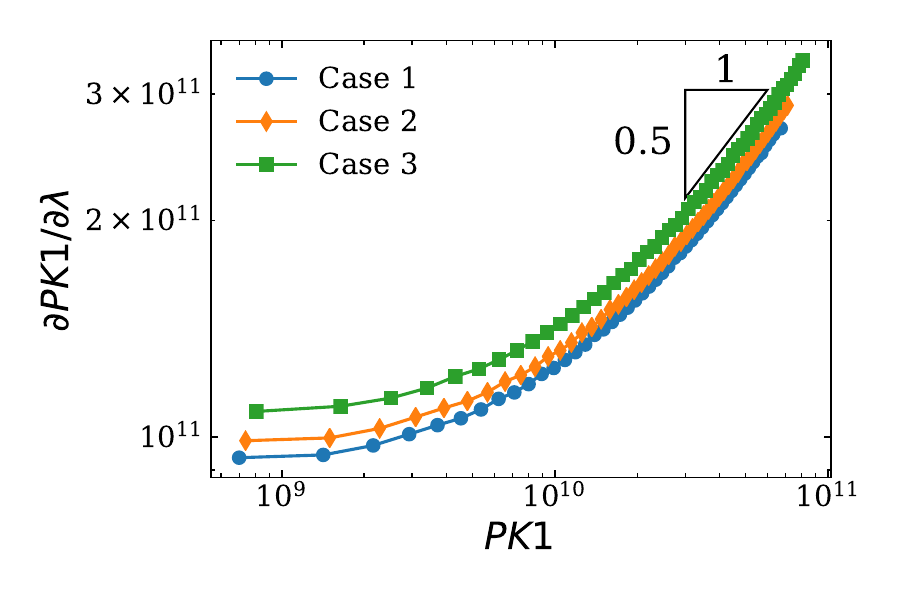}
        \caption{Tangent stiffness vs stress curve of FCL with various initial orientation fields.}
        \label{fig:tangent_stiffness}
    \end{subfigure}
    \caption{Homogenized mechanics of the FCL with various degrees of fiber alignment.}
    \label{fig:macro_mechanics}
\end{figure}

To this end, figure \ref{fig:macro_mechanics} shows the homogenized mechanics of the FCL undergoing uniaxial loading. The stress-strain curve shown in figure \ref{fig:stress_strain} indicates that the degree of alignment affects the overall mechanics of the FCL. Here, the three cases correspond to the distribution of RVE alignments shown in figure \ref{fig:degree_of_alignment}.

In addition, figure \ref{fig:tangent_stiffness} shows how the tangent stiffness varies with stress for different degrees of alignment. The curves reveal a gradual transition from an initial small strain stiffness (left end of curves in figure \ref{fig:tangent_stiffness}) to a power law relation between the stiffness and stress of exponent $1/2$. This exponent has been observed in discrete network models  and implies that the second Piola-Kirchhoff stress is proportional to the square of the Green-Lagrange strain \cite{zagarTwoFundamentalMechanisms2015}.

\section{Conclusion and Outlook} \label{sec:Conclusion}
\MuMFiM{} provides an open source tool for solving two-scale finite element problems with fibers at the microscale. The software architecture makes it easy to provide new material models on the micro and macro scales. By using multiple levels of parallelism, excellent strong and weak scalability have been achieved. Additionally, we have shown how many small RVEs can be simultaneously solved on the GPU for an up to 1000x performance improvement over a naive GPU-based approach. Current applications of \MuMFiM{} include modeling the FCL as well as other biological structures. These applications have led to multiple avenues of future research, which we hope to undertake. In the purely computational domain, we expect to implement dynamic load balancing and solution check-pointing in \MuMFiM{}. We are particularly interested in understanding how mesh adaption and error localization can couple with \MuMFiM{}'s current multiscale technology. Ideally, these strategies can be used to help reduce the computational cost of running a multiscale analysis by using large element sizes where there is limited problem physics occurring. Additionally, we believe that these will be the required first steps to understanding how to represent a microscale localization process on the macroscale mesh. We also hope to expand our macroscale field description to take advantage of nonlocal continuum field theories. Nonlocality of network materials has been shown to be of great importance for capturing the underlying behavior and associated size effects. In addition, using a nonlocal continuum theory as the macroscale continuum description allows for reduced scale separation, that is, the microscale RVE doesn't need to be an order of magnitude smaller than the macroscale mesh element. This is helpful to avoid a minimum macroscale element size and eases the implementation of automatic remeshing.

\section{Acknowledgements}
This work was supported in part by the National Institutes of
Health (NIH) through Grant No. U01 AT010326-06. This material
is based in part upon work supported by the National Science
Foundation Graduate Research Fellowship under Grant No. DGE-1744655. Computational resources were provided by the Rensselaer Polytechnic Institute Center for Computational Innovations that is partly supported by a National Science Foundation through MRI grant number 1828083. 

\section{Conflict of Interest}
All authors certify that they have no affiliations with or involvement in any organization or entity with any financial interest or non-financial interest in the subject matter or materials discussed in this manuscript.

\bibliography{MUMFIM-EWC,manual-additions}


\begin{thebibliography}{83}
\ifx \bisbn   \undefined \def \bisbn  #1{ISBN #1}\fi
\ifx \binits  \undefined \def \binits#1{#1}\fi
\ifx \bauthor  \undefined \def \bauthor#1{#1}\fi
\ifx \batitle  \undefined \def \batitle#1{#1}\fi
\ifx \bjtitle  \undefined \def \bjtitle#1{#1}\fi
\ifx \bvolume  \undefined \def \bvolume#1{\textbf{#1}}\fi
\ifx \byear  \undefined \def \byear#1{#1}\fi
\ifx \bissue  \undefined \def \bissue#1{#1}\fi
\ifx \bfpage  \undefined \def \bfpage#1{#1}\fi
\ifx \blpage  \undefined \def \blpage #1{#1}\fi
\ifx \burl  \undefined \def \burl#1{\textsf{#1}}\fi
\ifx \doiurl  \undefined \def \doiurl#1{\url{https://doi.org/#1}}\fi
\ifx \betal  \undefined \def \betal{\textit{et al.}}\fi
\ifx \binstitute  \undefined \def \binstitute#1{#1}\fi
\ifx \binstitutionaled  \undefined \def \binstitutionaled#1{#1}\fi
\ifx \bctitle  \undefined \def \bctitle#1{#1}\fi
\ifx \beditor  \undefined \def \beditor#1{#1}\fi
\ifx \bpublisher  \undefined \def \bpublisher#1{#1}\fi
\ifx \bbtitle  \undefined \def \bbtitle#1{#1}\fi
\ifx \bedition  \undefined \def \bedition#1{#1}\fi
\ifx \bseriesno  \undefined \def \bseriesno#1{#1}\fi
\ifx \blocation  \undefined \def \blocation#1{#1}\fi
\ifx \bsertitle  \undefined \def \bsertitle#1{#1}\fi
\ifx \bsnm \undefined \def \bsnm#1{#1}\fi
\ifx \bsuffix \undefined \def \bsuffix#1{#1}\fi
\ifx \bparticle \undefined \def \bparticle#1{#1}\fi
\ifx \barticle \undefined \def \barticle#1{#1}\fi
\bibcommenthead
\ifx \bconfdate \undefined \def \bconfdate #1{#1}\fi
\ifx \botherref \undefined \def \botherref #1{#1}\fi
\ifx \url \undefined \def \url#1{\textsf{#1}}\fi
\ifx \bchapter \undefined \def \bchapter#1{#1}\fi
\ifx \bbook \undefined \def \bbook#1{#1}\fi
\ifx \bcomment \undefined \def \bcomment#1{#1}\fi
\ifx \oauthor \undefined \def \oauthor#1{#1}\fi
\ifx \citeauthoryear \undefined \def \citeauthoryear#1{#1}\fi
\ifx \endbibitem  \undefined \def \endbibitem {}\fi
\ifx \bconflocation  \undefined \def \bconflocation#1{#1}\fi
\ifx \arxivurl  \undefined \def \arxivurl#1{\textsf{#1}}\fi
\csname PreBibitemsHook\endcsname

\bibitem{kanouteMultiscaleMethodsComposites2009}
\begin{barticle}
\bauthor{\bsnm{Kanout{\'e}}, \binits{P.}},
\bauthor{\bsnm{Boso}, \binits{D.P.}},
\bauthor{\bsnm{Chaboche}, \binits{J.L.}},
\bauthor{\bsnm{Schrefler}, \binits{B.A.}}:
\batitle{Multiscale methods for composites: A review}.
\bjtitle{Archives of Computational Methods in Engineering}
\bvolume{16}(\bissue{1}),
\bfpage{31}--\blpage{75}
(\byear{2009}).
\doiurl{10.1007/s11831-008-9028-8}
\end{barticle}
\endbibitem

\bibitem{feyelMultiscaleFE2Elastoviscoplastic1999}
\begin{barticle}
\bauthor{\bsnm{Feyel}, \binits{F.}}:
\batitle{Multiscale fe2 elastoviscoplastic analysis of composite structures}.
\bjtitle{Computational Materials Science}
\bvolume{16}(\bissue{1}),
\bfpage{344}--\blpage{354}
(\byear{1999}).
\doiurl{10.1016/S0927-0256(99)00077-4}
\end{barticle}
\endbibitem

\bibitem{feyelMultiscaleNonLinear2001}
\begin{barticle}
\bauthor{\bsnm{Feyel}, \binits{F.}}:
\batitle{Multiscale non linear fe2 analysis of composite structures : Fiber
  size effects}.
\bjtitle{Le Journal de Physique IV}
\bvolume{11}(\bissue{PR5}),
\bfpage{195}--\blpage{202}
(\byear{2001}).
\doiurl{10.1051/jp4:2001524}
\end{barticle}
\endbibitem

\bibitem{nezamabadiMultilevelComputationalStrategy2009}
\begin{barticle}
\bauthor{\bsnm{Nezamabadi}, \binits{S.}},
\bauthor{\bsnm{Yvonnet}, \binits{J.}},
\bauthor{\bsnm{Zahrouni}, \binits{H.}},
\bauthor{\bsnm{{Potier-Ferry}}, \binits{M.}}:
\batitle{A multilevel computational strategy for handling microscopic and
  macroscopic instabilities}.
\bjtitle{Computer Methods in Applied Mechanics and Engineering}
\bvolume{198}(\bissue{27-29}),
\bfpage{2099}--\blpage{2110}
(\byear{2009}).
\doiurl{10.1016/j.cma.2009.02.026}
\end{barticle}
\endbibitem

\bibitem{tikarrouchineThreedimensionalFE2Method2018}
\begin{barticle}
\bauthor{\bsnm{Tikarrouchine}, \binits{E.}},
\bauthor{\bsnm{Chatzigeorgiou}, \binits{G.}},
\bauthor{\bsnm{Praud}, \binits{F.}},
\bauthor{\bsnm{Piotrowski}, \binits{B.}},
\bauthor{\bsnm{Chemisky}, \binits{Y.}},
\bauthor{\bsnm{Meraghni}, \binits{F.}}:
\batitle{Three-dimensional fe2 method for the simulation of non-linear,
  rate-dependent response of composite structures}.
\bjtitle{Composite Structures}
\bvolume{193},
\bfpage{165}--\blpage{179}
(\byear{2018}).
\doiurl{10.1016/j.compstruct.2018.03.072}
\end{barticle}
\endbibitem

\bibitem{eidelHeterogeneousMultiscaleFinite2018}
\begin{barticle}
\bauthor{\bsnm{Eidel}, \binits{B.}},
\bauthor{\bsnm{Fischer}, \binits{A.}}:
\batitle{The heterogeneous multiscale finite element method for the
  homogenization of linear elastic solids and a comparison with the fe2
  method}.
\bjtitle{Computer Methods in Applied Mechanics and Engineering}
\bvolume{329},
\bfpage{332}--\blpage{368}
(\byear{2018}).
\doiurl{10.1016/j.cma.2017.10.001}
\end{barticle}
\endbibitem

\bibitem{mieheComputationalMicrotomacroTransitions2003}
\begin{barticle}
\bauthor{\bsnm{Miehe}, \binits{C.}}:
\batitle{Computational micro-to-macro transitions for discretized
  micro-structures of heterogeneous materials at finite strains based on the
  minimization of averaged incremental energy}.
\bjtitle{Computer Methods in Applied Mechanics and Engineering}
\bvolume{192}(\bissue{5-6}),
\bfpage{559}--\blpage{591}
(\byear{2003}).
\doiurl{10.1016/S0045-7825(02)00564-9}
\end{barticle}
\endbibitem

\bibitem{smitPredictionMechanicalBehavior1998}
\begin{barticle}
\bauthor{\bsnm{Smit}, \binits{R.J.M.}},
\bauthor{\bsnm{Brekelmans}, \binits{W.A.M.}},
\bauthor{\bsnm{Meijer}, \binits{H.E.H.}}:
\batitle{Prediction of the mechanical behavior of nonlinear heterogeneous
  systems by multi-level finite element modeling}.
\bjtitle{Computer Methods in Applied Mechanics and Engineering}
\bvolume{155}(\bissue{1-2}),
\bfpage{181}--\blpage{192}
(\byear{1998}).
\doiurl{10.1016/S0045-7825(97)00139-4}
\end{barticle}
\endbibitem

\bibitem{feyelMultiscaleNonlinearFE22001}
\begin{barticle}
\bauthor{\bsnm{Feyel}, \binits{F.}},
\bauthor{\bsnm{Chaboche}, \binits{J.-L.}}:
\batitle{Multi-scale non-linear fe2 analysis of composite structures: Damage
  and fiber size effects}.
\bjtitle{Revue Europ\'eenne des \'El\'ements Finis}
\bvolume{10}(\bissue{2-4}),
\bfpage{449}--\blpage{472}
(\byear{2001}).
\doiurl{10.1080/12506559.2001.11869262}
\end{barticle}
\endbibitem

\bibitem{feyelMultilevelFiniteElement2003}
\begin{barticle}
\bauthor{\bsnm{Feyel}, \binits{F.}}:
\batitle{A multilevel finite element method (fe2) to describe the response of
  highly non-linear structures using generalized continua}.
\bjtitle{Computer Methods in Applied Mechanics and Engineering}
\bvolume{192}(\bissue{28-30}),
\bfpage{3233}--\blpage{3244}
(\byear{2003}).
\doiurl{10.1016/S0045-7825(03)00348-7}
\end{barticle}
\endbibitem

\bibitem{castanedaEffectiveMechanicalProperties1991}
\begin{barticle}
\bauthor{\bsnm{Casta{\~n}eda}, \binits{P.P.}}:
\batitle{The effective mechanical properties of nonlinear isotropic
  composites}.
\bjtitle{Journal of the Mechanics and Physics of Solids}
\bvolume{39}(\bissue{1}),
\bfpage{45}--\blpage{71}
(\byear{1991}).
\doiurl{10.1016/0022-5096(91)90030-R}
\end{barticle}
\endbibitem

\bibitem{moulinecFastNumericalMethod1994}
\begin{botherref}
\oauthor{\bsnm{Moulinec}, \binits{H.}},
\oauthor{\bsnm{Suquet}, \binits{P.}}:
A fast numerical method for computing the linear and nonlinear mechanical
  properties of composites.
Comptes Rendus de l'Acad\'emie des sciences. S\'erie II. M\'ecanique, physique,
  chimie, astronomie
(1994)
\end{botherref}
\endbibitem

\bibitem{moulinecNumericalMethodComputing1998}
\begin{barticle}
\bauthor{\bsnm{Moulinec}, \binits{H.}},
\bauthor{\bsnm{Suquet}, \binits{P.}}:
\batitle{A numerical method for computing the overall response of nonlinear
  composites with complex microstructure}.
\bjtitle{Computer Methods in Applied Mechanics and Engineering}
\bvolume{157}(\bissue{1-2}),
\bfpage{69}--\blpage{94}
(\byear{1998})
{\href{https://arxiv.org/abs/2012.08962}{{arxiv:2012.08962}}}
{[physics]}.
\doiurl{10.1016/S0045-7825(97)00218-1}
\end{barticle}
\endbibitem

\bibitem{michelEffectivePropertiesComposite1999}
\begin{barticle}
\bauthor{\bsnm{Michel}, \binits{J.C.}},
\bauthor{\bsnm{Moulinec}, \binits{H.}},
\bauthor{\bsnm{Suquet}, \binits{P.}}:
\batitle{Effective properties of composite materials with periodic
  microstructure: A computational approach}.
\bjtitle{Computer Methods in Applied Mechanics and Engineering}
\bvolume{172}(\bissue{1-4}),
\bfpage{109}--\blpage{143}
(\byear{1999}).
\doiurl{10.1016/S0045-7825(98)00227-8}
\end{barticle}
\endbibitem

\bibitem{geersHomogenizationMethodsMultiscale2017}
\begin{bchapter}
\bauthor{\bsnm{Geers}, \binits{M.G.D.}},
\bauthor{\bsnm{Kouznetsova}, \binits{V.G.}},
\bauthor{\bsnm{Matou{\v s}}, \binits{K.}},
\bauthor{\bsnm{Yvonnet}, \binits{J.}}:
\bctitle{Homogenization methods and multiscale modeling: Nonlinear problems}.
In: \bbtitle{Encyclopedia of Computational Mechanics Second Edition},
pp. \bfpage{1}--\blpage{34}.
\bpublisher{John Wiley \& Sons, Ltd}, \blocation{???}
(\byear{2017}).
\doiurl{10.1002/9781119176817.ecm2107}
\end{bchapter}
\endbibitem

\bibitem{kouznetsovaApproachMicromacroModeling2001}
\begin{barticle}
\bauthor{\bsnm{Kouznetsova}, \binits{V.}},
\bauthor{\bsnm{Brekelmans}, \binits{W.A.M.}},
\bauthor{\bsnm{Baaijens}, \binits{F.P.T.}}:
\batitle{An approach to micro-macro modeling of heterogeneous materials}.
\bjtitle{Computational Mechanics}
\bvolume{27}(\bissue{1}),
\bfpage{37}--\blpage{48}
(\byear{2001}).
\doiurl{10.1007/s004660000212}
\end{barticle}
\endbibitem

\bibitem{coenenMultiscaleApproachBridge2012}
\begin{barticle}
\bauthor{\bsnm{Coenen}, \binits{E.W.C.}},
\bauthor{\bsnm{Kouznetsova}, \binits{V.G.}},
\bauthor{\bsnm{Bosco}, \binits{E.}},
\bauthor{\bsnm{Geers}, \binits{M.G.D.}}:
\batitle{A multi-scale approach to bridge microscale damage and macroscale
  failure: A nested computational homogenization-localization framework}.
\bjtitle{International Journal of Fracture}
\bvolume{178}(\bissue{1-2}),
\bfpage{157}--\blpage{178}
(\byear{2012}).
\doiurl{10.1007/s10704-012-9765-4}
\end{barticle}
\endbibitem

\bibitem{pontecastanedaVariationalLinearComparison2023}
\begin{barticle}
\bauthor{\bsnm{Ponte~Casta{\~n}eda}, \binits{P.}}:
\batitle{Variational linear comparison homogenization estimates for the flow of
  yield stress fluids through porous media}.
\bjtitle{Journal of Non-Newtonian Fluid Mechanics}
\bvolume{321},
\bfpage{105104}
(\byear{2023}).
\doiurl{10.1016/j.jnnfm.2023.105104}
\end{barticle}
\endbibitem

\bibitem{kouznetsovaMultiscaleSecondorderComputational2004}
\begin{barticle}
\bauthor{\bsnm{Kouznetsova}, \binits{V.G.}},
\bauthor{\bsnm{Geers}, \binits{M.G.D.}},
\bauthor{\bsnm{Brekelmans}, \binits{W.A.M.}}:
\batitle{Multi-scale second-order computational homogenization of multi-phase
  materials: A nested finite element solution strategy}.
\bjtitle{Computer Methods in Applied Mechanics and Engineering}
\bvolume{193}(\bissue{48-51}),
\bfpage{5525}--\blpage{5550}
(\byear{2004}).
\doiurl{10.1016/j.cma.2003.12.073}
\end{barticle}
\endbibitem

\bibitem{temizerNumericalMethodHomogenization2007}
\begin{barticle}
\bauthor{\bsnm{Temizer}, \binits{I.}},
\bauthor{\bsnm{Zohdi}, \binits{T.I.}}:
\batitle{A numerical method for homogenization in non-linear elasticity}.
\bjtitle{Computational Mechanics}
\bvolume{40}(\bissue{2}),
\bfpage{281}--\blpage{298}
(\byear{2007}).
\doiurl{10.1007/s00466-006-0097-y}
\end{barticle}
\endbibitem

\bibitem{zhiDirectFEModeling2022}
\begin{barticle}
\bauthor{\bsnm{Zhi}, \binits{J.}},
\bauthor{\bsnm{Poh}, \binits{L.H.}},
\bauthor{\bsnm{Tay}, \binits{T.-E.}},
\bauthor{\bsnm{Tan}, \binits{V.B.C.}}:
\batitle{Direct fe 2 modeling of heterogeneous materials with a micromorphic
  computational homogenization framework}.
\bjtitle{Computer Methods in Applied Mechanics and Engineering}
\bvolume{393},
\bfpage{114837}
(\byear{2022}).
\doiurl{10.1016/j.cma.2022.114837}
\end{barticle}
\endbibitem

\bibitem{temizerComputationMacroscopicTangent2008}
\begin{barticle}
\bauthor{\bsnm{Temizer}, \binits{{\. I}.}},
\bauthor{\bsnm{Wriggers}, \binits{P.}}:
\batitle{On the computation of the macroscopic tangent for multiscale
  volumetric homogenization problems}.
\bjtitle{Computer Methods in Applied Mechanics and Engineering}
\bvolume{198}(\bissue{3-4}),
\bfpage{495}--\blpage{510}
(\byear{2008}).
\doiurl{10.1016/j.cma.2008.08.018}
\end{barticle}
\endbibitem

\bibitem{ibanezPUMIParallelUnstructured2016}
\begin{barticle}
\bauthor{\bsnm{Ibanez}, \binits{D.A.}},
\bauthor{\bsnm{Seol}, \binits{E.S.}},
\bauthor{\bsnm{Smith}, \binits{C.W.}},
\bauthor{\bsnm{Shephard}, \binits{M.S.}}:
\batitle{Pumi: Parallel unstructured mesh infrastructure}.
\bjtitle{ACM Transactions on Mathematical Software}
\bvolume{42}(\bissue{3}),
\bfpage{1}--\blpage{28}
(\byear{2016}).
\doiurl{10.1145/2814935}
\end{barticle}
\endbibitem

\bibitem{balayEfficientManagementParallelism1997}
\begin{bchapter}
\bauthor{\bsnm{Balay}, \binits{S.}},
\bauthor{\bsnm{Gropp}, \binits{W.D.}},
\bauthor{\bsnm{McInnes}, \binits{L.C.}},
\bauthor{\bsnm{Smith}, \binits{B.F.}}:
\bctitle{Efficient management of parallelism in object oriented numerical
  software libraries}.
In: \beditor{\bsnm{Arge}, \binits{E.}},
\beditor{\bsnm{Bruaset}, \binits{A.M.}},
\beditor{\bsnm{Langtangen}, \binits{H.P.}} (eds.)
\bbtitle{Modern Aoftware Tools in Scientific Computing},
pp. \bfpage{163}--\blpage{202}.
\bpublisher{Birkh\"auser Press},
\blocation{Basel, Switzerland}
(\byear{1997})
\end{bchapter}
\endbibitem

\bibitem{balayPETScUsersManual2018}
\begin{botherref}
\oauthor{\bsnm{Balay}, \binits{S.}},
\oauthor{\bsnm{Abhyankar}, \binits{S.}},
\oauthor{\bsnm{Adams}, \binits{M.F.}},
\oauthor{\bsnm{Brown}, \binits{J.}},
\oauthor{\bsnm{Brune}, \binits{P.}},
\oauthor{\bsnm{Buschelman}, \binits{K.}},
\oauthor{\bsnm{Dalcin}, \binits{L.}},
\oauthor{\bsnm{Dener}, \binits{A.}},
\oauthor{\bsnm{Eijkhout}, \binits{V.}},
\oauthor{\bsnm{Gropp}, \binits{W.D.}},
\oauthor{\bsnm{Kaushik}, \binits{D.}},
\oauthor{\bsnm{Knepley}, \binits{M.G.}},
\oauthor{\bsnm{May}, \binits{D.A.}},
\oauthor{\bsnm{McInnes}, \binits{L.C.}},
\oauthor{\bsnm{Mills}, \binits{R.T.}},
\oauthor{\bsnm{Munson}, \binits{T.}},
\oauthor{\bsnm{Rupp}, \binits{K.}},
\oauthor{\bsnm{Sanan}, \binits{P.}},
\oauthor{\bsnm{Smith}, \binits{B.F.}},
\oauthor{\bsnm{Zampini}, \binits{S.}},
\oauthor{\bsnm{Zhang}, \binits{H.}},
\oauthor{\bsnm{Zhang}, \binits{H.}}:
Petsc users manual.
Manual,
Argonne National Laboratory
(2018)
\end{botherref}
\endbibitem

\bibitem{underwoodDynamicRelaxation1986}
\begin{bchapter}
\bauthor{\bsnm{Underwood}, \binits{P.}}:
\bctitle{Dynamic relaxation}.
In: \beditor{\bsnm{Belytschko}, \binits{T.}},
\beditor{\bsnm{Hughes}, \binits{T.J.R.}} (eds.)
\bbtitle{Computational Method for Transient Analysis}.
\bsertitle{Computational Methods in Mechanics},
vol. \bseriesno{1},
pp. \bfpage{245}--\blpage{263}.
\bpublisher{North Holland},
\blocation{Amsterdam, Denmark}
(\byear{1986})
\end{bchapter}
\endbibitem

\bibitem{mahutgaNonaffineFiberNetwork2023}
\begin{barticle}
\bauthor{\bsnm{Mahutga}, \binits{R.R.}},
\bauthor{\bsnm{Barocas}, \binits{V.H.}},
\bauthor{\bsnm{Alford}, \binits{P.W.}}:
\batitle{The non-affine fiber network solver: A multiscale fiber network
  material model for finite-element analysis}.
\bjtitle{Journal of the Mechanical Behavior of Biomedical Materials}
\bvolume{144},
\bfpage{105967}
(\byear{2023}).
\doiurl{10.1016/j.jmbbm.2023.105967}
\end{barticle}
\endbibitem

\bibitem{glugeGeneralizedBoundaryConditions2013}
\begin{barticle}
\bauthor{\bsnm{Gl{\"u}ge}, \binits{R.}}:
\batitle{Generalized boundary conditions on representative volume elements and
  their use in determining the effective material properties}.
\bjtitle{Computational Materials Science}
\bvolume{79},
\bfpage{408}--\blpage{416}
(\byear{2013}).
\doiurl{10.1016/j.commatsci.2013.06.038}
\end{barticle}
\endbibitem

\bibitem{hillElasticPropertiesReinforced1963}
\begin{barticle}
\bauthor{\bsnm{Hill}, \binits{R.}}:
\batitle{Elastic properties of reinforced solids: Some theoretical principles}.
\bjtitle{Journal of the Mechanics and Physics of Solids}
\bvolume{11}(\bissue{5}),
\bfpage{357}--\blpage{372}
(\byear{1963}).
\doiurl{10.1016/0022-5096(63)90036-X}
\end{barticle}
\endbibitem

\bibitem{liuDiscreteAveragingRelations2016}
\begin{barticle}
\bauthor{\bsnm{Liu}, \binits{C.}},
\bauthor{\bsnm{Reina}, \binits{C.}}:
\batitle{Discrete averaging relations for micro to macro transition}.
\bjtitle{Journal of Applied Mechanics}
\bvolume{83}(\bissue{8}),
\bfpage{081006}
(\byear{2016}).
\doiurl{10.1115/1.4033552}
\end{barticle}
\endbibitem

\bibitem{walters2021considering}
\begin{barticle}
\bauthor{\bsnm{Walters}, \binits{D.J.}},
\bauthor{\bsnm{Luscher}, \binits{D.J.}},
\bauthor{\bsnm{Yeager}, \binits{J.D.}}:
\batitle{Considering computational speed vs. accuracy: Choosing appropriate
  mesoscale rve boundary conditions}.
\bjtitle{Computer Methods in Applied Mechanics and Engineering}
\bvolume{374},
\bfpage{113572}
(\byear{2021})
\end{barticle}
\endbibitem

\bibitem{luscher2012essential}
\begin{botherref}
\oauthor{\bsnm{Luscher}, \binits{D.}},
\oauthor{\bsnm{McDowell}, \binits{D.L.}},
\oauthor{\bsnm{Bronkhorst}, \binits{C.}}:
Essential features of fine scale boundary conditions for second gradient
  multiscale homogenization of statistical volume elements.
International Journal for Multiscale Computational Engineering
\textbf{10}(5)
(2012)
\end{botherref}
\endbibitem

\bibitem{mersonSizeEffectsRandom2020}
\begin{barticle}
\bauthor{\bsnm{Merson}, \binits{J.}},
\bauthor{\bsnm{Picu}, \binits{R.C.}}:
\batitle{Size effects in random fiber networks controlled by the use of
  generalized boundary conditions}.
\bjtitle{International Journal of Solids and Structures}
\bvolume{206},
\bfpage{314}--\blpage{321}
(\byear{2020}).
\doiurl{10.1016/j.ijsolstr.2020.09.033}
\end{barticle}
\endbibitem

\bibitem{shahsavariSizeEffectMechanical2013}
\begin{barticle}
\bauthor{\bsnm{Shahsavari}, \binits{A.S.}},
\bauthor{\bsnm{Picu}, \binits{R.C.}}:
\batitle{Size effect on mechanical behavior of random fiber networks}.
\bjtitle{International Journal of Solids and Structures}
\bvolume{50}(\bissue{20}),
\bfpage{3332}--\blpage{3338}
(\byear{2013}).
\doiurl{10.1016/j.ijsolstr.2013.06.004}
\end{barticle}
\endbibitem

\bibitem{licupElasticRegimesSubisostatic2016}
\begin{barticle}
\bauthor{\bsnm{Licup}, \binits{A.J.}},
\bauthor{\bsnm{Sharma}, \binits{A.}},
\bauthor{\bsnm{MacKintosh}, \binits{F.C.}}:
\batitle{Elastic regimes of subisostatic athermal fiber networks}.
\bjtitle{Physical Review E}
\bvolume{93}(\bissue{1}),
\bfpage{012407}
(\byear{2016}).
\doiurl{10.1103/PhysRevE.93.012407}
\end{barticle}
\endbibitem

\bibitem{islamStochasticContinuumModel2018}
\begin{barticle}
\bauthor{\bsnm{Islam}, \binits{M.R.}},
\bauthor{\bsnm{Tudryn}, \binits{G.}},
\bauthor{\bsnm{Bucinell}, \binits{R.}},
\bauthor{\bsnm{Schadler}, \binits{L.}},
\bauthor{\bsnm{Picu}, \binits{R.C.}}:
\batitle{Stochastic continuum model for mycelium-based bio-foam}.
\bjtitle{Materials and Design}
\bvolume{160},
\bfpage{549}--\blpage{556}
(\byear{2018}).
\doiurl{10.1016/j.matdes.2018.09.046}
\end{barticle}
\endbibitem

\bibitem{deogekarStrengthRandomFiber2018}
\begin{barticle}
\bauthor{\bsnm{Deogekar}, \binits{S.}},
\bauthor{\bsnm{Picu}, \binits{R.C.}}:
\batitle{On the strength of random fiber networks}.
\bjtitle{Journal of the Mechanics and Physics of Solids}
\bvolume{116},
\bfpage{1}--\blpage{16}
(\byear{2018}).
\doiurl{10.1016/j.jmps.2018.03.026}
\end{barticle}
\endbibitem

\bibitem{courantUberPartiellenDifferenzengleichungen1928}
\begin{barticle}
\bauthor{\bsnm{Courant}, \binits{R.}},
\bauthor{\bsnm{Friedrichs}, \binits{K.}},
\bauthor{\bsnm{Lewy}, \binits{H.}}:
\batitle{\"uber die partiellen differenzengleichungen der mathematischen
  physik}.
\bjtitle{Mathematische Annalen}
\bvolume{100}(\bissue{1}),
\bfpage{32}--\blpage{74}
(\byear{1928}).
\doiurl{10.1007/BF01448839}
\end{barticle}
\endbibitem

\bibitem{belytschkoNonlinearFiniteElements2014}
\begin{bbook}
\bauthor{\bsnm{Belytschko}, \binits{T.}},
\bauthor{\bsnm{Liu}, \binits{W.K.}},
\bauthor{\bsnm{Moran}, \binits{B.}},
\bauthor{\bsnm{Elkhodary}, \binits{K.I.}}:
\bbtitle{Nonlinear Finite Elements for Continua and Structures},
\bedition{2nd} edn.
\bpublisher{Wiley},
\blocation{Chichester, West Sussex, U.K.}
(\byear{2014})
\end{bbook}
\endbibitem

\bibitem{papadrakakisMethodAutomaticEvaluation1981}
\begin{barticle}
\bauthor{\bsnm{Papadrakakis}, \binits{M.}}:
\batitle{A method for the automatic evaluation of the dynamic relaxation
  parameters}.
\bjtitle{Computer Methods in Applied Mechanics and Engineering}
\bvolume{25}(\bissue{1}),
\bfpage{35}--\blpage{48}
(\byear{1981}).
\doiurl{10.1016/0045-7825(81)90066-9}
\end{barticle}
\endbibitem

\bibitem{zhangDevelopmentMaDRMethod1994}
\begin{barticle}
\bauthor{\bsnm{Zhang}, \binits{L.C.}},
\bauthor{\bsnm{Kadkhodayan}, \binits{M.}},
\bauthor{\bsnm{Mai}, \binits{Y.-W.}}:
\batitle{Development of the madr method}.
\bjtitle{Computers and Structures}
\bvolume{52}(\bissue{1}),
\bfpage{1}--\blpage{8}
(\byear{1994}).
\doiurl{10.1016/0045-7949(94)90249-6}
\end{barticle}
\endbibitem

\bibitem{rezaiee-pajandDynamicRelaxationMethod2010}
\begin{barticle}
\bauthor{\bsnm{{Rezaiee-Pajand}}, \binits{M.}},
\bauthor{\bsnm{Alamatian}, \binits{J.}}:
\batitle{The dynamic relaxation method using new formulation for fictitious
  mass and damping}.
\bjtitle{Structural Engineering and Mechanics}
\bvolume{34}(\bissue{1}),
\bfpage{109}--\blpage{133}
(\byear{2010}).
\doiurl{10.12989/SEM.2010.34.1.109}
\end{barticle}
\endbibitem

\bibitem{leeSimpleExplicitArclength2011}
\begin{barticle}
\bauthor{\bsnm{Lee}, \binits{K.S.}},
\bauthor{\bsnm{Han}, \binits{S.E.}},
\bauthor{\bsnm{Park}, \binits{T.}}:
\batitle{A simple explicit arc-length method using the dynamic relaxation
  method with kinetic damping}.
\bjtitle{Computers and Structures}
\bvolume{89}(\bissue{1-2}),
\bfpage{216}--\blpage{233}
(\byear{2011}).
\doiurl{10.1016/j.compstruc.2010.09.006}
\end{barticle}
\endbibitem

\bibitem{gastonMOOSEParallelComputational2009}
\begin{barticle}
\bauthor{\bsnm{Gaston}, \binits{D.}},
\bauthor{\bsnm{Newman}, \binits{C.}},
\bauthor{\bsnm{Hansen}, \binits{G.}},
\bauthor{\bsnm{{Lebrun-Grandie}}, \binits{D.}}:
\batitle{Moose: A parallel computational framework for coupled systems of
  nonlinear equations}.
\bjtitle{Nuclear Engineering and Design}
\bvolume{239}(\bissue{10}),
\bfpage{1768}--\blpage{1778}
(\byear{2009})
\end{barticle}
\endbibitem

\bibitem{davisondestgermainUintahMassivelyParallel2000}
\begin{bchapter}
\bauthor{\bsnm{{Davison de St Germain}}, \binits{J.}},
\bauthor{\bsnm{McCorquodale}, \binits{J.}},
\bauthor{\bsnm{Parker}, \binits{S.G.}},
\bauthor{\bsnm{Johnson}, \binits{C.R.}}:
\bctitle{Uintah: A massively parallel problem solving environment}.
In: \bbtitle{Proceedings the Ninth International Symposium on High-Performance
  Distributed Computing},
pp. \bfpage{33}--\blpage{41}.
\bpublisher{IEEE}, \blocation{???}
(\byear{2000})
\end{bchapter}
\endbibitem

\bibitem{joppichMpCCIToolSimulation2006}
\begin{barticle}
\bauthor{\bsnm{Joppich}, \binits{W.}},
\bauthor{\bsnm{K{\"u}rschner}, \binits{M.}}:
\batitle{Mpcci\textemdash a tool for the simulation of coupled applications}.
\bjtitle{Concurrency and Computation: Practice and Experience}
\bvolume{18}(\bissue{2}),
\bfpage{183}--\blpage{192}
(\byear{2006}).
\doiurl{10.1002/cpe.913}
\end{barticle}
\endbibitem

\bibitem{gravemeierTaxonomyMultiscaleMethods2008}
\begin{barticle}
\bauthor{\bsnm{Gravemeier}, \binits{V.}},
\bauthor{\bsnm{Lenz}, \binits{S.}},
\bauthor{\bsnm{Wall}, \binits{W.A.}}:
\batitle{Towards a taxonomy for multiscale methods in computational mechanics:
  Building blocks of existing methods}.
\bjtitle{Computational Mechanics}
\bvolume{41}(\bissue{2}),
\bfpage{279}--\blpage{291}
(\byear{2008})
\end{barticle}
\endbibitem

\bibitem{tobinAdaptiveMultiscaleSimulation2018}
\begin{botherref}
\oauthor{\bsnm{Tobin}, \binits{W.R.}}:
The adaptive multiscale simulation infrastructure.
PhD thesis,
Rensselaer Polytechn. Inst.,
Troy, NY USA
(2018).
Dept. Comput. Sci.
\end{botherref}
\endbibitem

\bibitem{mersonModeltraitsModelAttribute2021}
\begin{barticle}
\bauthor{\bsnm{Merson}, \binits{J.}},
\bauthor{\bsnm{Shephard}, \binits{M.S.}}:
\batitle{Model-traits: Model attribute definitions for scientific simulations
  in c++}.
\bjtitle{The Journal of Open Source Software}
\bvolume{6}(\bissue{64}),
\bfpage{3389}
(\byear{2021}).
\doiurl{10.21105/joss.03389}
\end{barticle}
\endbibitem

\bibitem{carteredwardsKokkosEnablingManycore2014}
\begin{barticle}
\bauthor{\bsnm{Carter~Edwards}, \binits{H.}},
\bauthor{\bsnm{Trott}, \binits{C.R.}},
\bauthor{\bsnm{Sunderland}, \binits{D.}}:
\batitle{Kokkos: Enabling manycore performance portability through polymorphic
  memory access patterns}.
\bjtitle{Journal of Parallel and Distributed Computing}
\bvolume{74}(\bissue{12}),
\bfpage{3202}--\blpage{3216}
(\byear{2014}).
\doiurl{10.1016/j.jpdc.2014.07.003}
\end{barticle}
\endbibitem

\bibitem{trottKokkosProgrammingModel2022}
\begin{barticle}
\bauthor{\bsnm{Trott}, \binits{C.R.}},
\bauthor{\bsnm{{Lebrun-Grandi{\'e}}}, \binits{D.}},
\bauthor{\bsnm{Arndt}, \binits{D.}},
\bauthor{\bsnm{Ciesko}, \binits{J.}},
\bauthor{\bsnm{Dang}, \binits{V.}},
\bauthor{\bsnm{Ellingwood}, \binits{N.}},
\bauthor{\bsnm{Gayatri}, \binits{R.}},
\bauthor{\bsnm{Harvey}, \binits{E.}},
\bauthor{\bsnm{Hollman}, \binits{D.S.}},
\bauthor{\bsnm{Ibanez}, \binits{D.}},
\bauthor{\bsnm{Liber}, \binits{N.}},
\bauthor{\bsnm{Madsen}, \binits{J.}},
\bauthor{\bsnm{Miles}, \binits{J.}},
\bauthor{\bsnm{Poliakoff}, \binits{D.}},
\bauthor{\bsnm{Powell}, \binits{A.}},
\bauthor{\bsnm{Rajamanickam}, \binits{S.}},
\bauthor{\bsnm{Simberg}, \binits{M.}},
\bauthor{\bsnm{Sunderland}, \binits{D.}},
\bauthor{\bsnm{Turcksin}, \binits{B.}},
\bauthor{\bsnm{Wilke}, \binits{J.}}:
\batitle{Kokkos 3: Programming model extensions for the exascale era}.
\bjtitle{IEEE Transactions on Parallel and Distributed Systems}
\bvolume{33}(\bissue{4}),
\bfpage{805}--\blpage{817}
(\byear{2022}).
\doiurl{10.1109/TPDS.2021.3097283}
\end{barticle}
\endbibitem

\bibitem{korchAcceleratingExplicitODE2018}
\begin{barticle}
\bauthor{\bsnm{Korch}, \binits{M.}},
\bauthor{\bsnm{Werner}, \binits{T.}}:
\batitle{Accelerating explicit ode methods on gpus by kernel fusion}.
\bjtitle{Concurrency and Computation: Practice and Experience}
\bvolume{30}(\bissue{18}),
\bfpage{4470}
(\byear{2018}).
\doiurl{10.1002/cpe.4470}
\end{barticle}
\endbibitem

\bibitem{wangKernelFusionEffective2010}
\begin{bchapter}
\bauthor{\bsnm{Wang}, \binits{G.}},
\bauthor{\bsnm{Lin}, \binits{Y.}},
\bauthor{\bsnm{Yi}, \binits{W.}}:
\bctitle{Kernel fusion: An effective method for better power efficiency on
  multithreaded gpu}.
In: \bbtitle{2010 IEEE/ACM Int'l Conference on Green Computing and
  Communications Int'l Conference on Cyber, Physical and Social Computing},
pp. \bfpage{344}--\blpage{350}.
\bpublisher{IEEE},
\blocation{Hangzhou}
(\byear{2010}).
\doiurl{10.1109/GreenCom-CPSCom.2010.102}
\end{bchapter}
\endbibitem

\bibitem{wahibScalableKernelFusion2014}
\begin{bchapter}
\bauthor{\bsnm{Wahib}, \binits{M.}},
\bauthor{\bsnm{Maruyama}, \binits{N.}}:
\bctitle{Scalable kernel fusion for memory-bound gpu applications}.
In: \bbtitle{Proceedings of the International Conference for High Performance
  Computing, Networking, Storage and Analysis}.
\bsertitle{SC '14},
pp. \bfpage{191}--\blpage{202}.
\bpublisher{IEEE Press},
\blocation{New Orleans, Louisana}
(\byear{2014}).
\doiurl{10.1109/SC.2014.21}
\end{bchapter}
\endbibitem

\bibitem{mersonKokkospackeddata2020}
\begin{botherref}
\oauthor{\bsnm{Merson}, \binits{J.}}:
Kokkos-Packed-Data
(2020)
\end{botherref}
\endbibitem

\bibitem{corralesImportanceCervicalCapsular2021}
\begin{barticle}
\bauthor{\bsnm{Corrales}, \binits{M.A.}},
\bauthor{\bsnm{Cronin}, \binits{D.S.}}:
\batitle{Importance of the cervical capsular joint cartilage geometry on head
  and facet joint kinematics assessed in a finite element neck model}.
\bjtitle{Journal of Biomechanics}
\bvolume{123},
\bfpage{110528}
(\byear{2021}).
\doiurl{10.1016/j.jbiomech.2021.110528}
\end{barticle}
\endbibitem

\bibitem{bermelAsymmetricInplaneShear2020}
\begin{barticle}
\bauthor{\bsnm{Bermel}, \binits{E.A.}},
\bauthor{\bsnm{Thakral}, \binits{S.}},
\bauthor{\bsnm{Claeson}, \binits{A.A.}},
\bauthor{\bsnm{Ellingson}, \binits{A.M.}},
\bauthor{\bsnm{Barocas}, \binits{V.H.}}:
\batitle{Asymmetric in-plane shear behavior of isolated cadaveric lumbar facet
  capsular ligaments: Implications for subject specific biomechanical models}.
\bjtitle{Journal of Biomechanics}
\bvolume{105},
\bfpage{109814}
(\byear{2020}).
\doiurl{10.1016/j.jbiomech.2020.109814}
\end{barticle}
\endbibitem

\bibitem{gacekThroughthicknessRegionalVariation2021}
\begin{barticle}
\bauthor{\bsnm{Gacek}, \binits{E.}},
\bauthor{\bsnm{Bermel}, \binits{E.A.}},
\bauthor{\bsnm{Ellingson}, \binits{A.M.}},
\bauthor{\bsnm{Barocas}, \binits{V.H.}}:
\batitle{Through-thickness regional variation in the mechanical characteristics
  of the lumbar facet capsular ligament}.
\bjtitle{Biomechanics and Modeling in Mechanobiology}
\bvolume{20}(\bissue{4}),
\bfpage{1445}--\blpage{1457}
(\byear{2021}).
\doiurl{10.1007/s10237-021-01455-3}
\end{barticle}
\endbibitem

\bibitem{itaIntraarticularCollagenaseSpinal2020}
\begin{barticle}
\bauthor{\bsnm{Ita}, \binits{M.E.}},
\bauthor{\bsnm{Ghimire}, \binits{P.}},
\bauthor{\bsnm{Welch}, \binits{R.L.}},
\bauthor{\bsnm{Troche}, \binits{H.R.}},
\bauthor{\bsnm{Winkelstein}, \binits{B.A.}}:
\batitle{Intra-articular collagenase in the spinal facet joint induces pain,
  drg neuron dysregulation and increased mmp-1 absent evidence of joint
  destruction}.
\bjtitle{Scientific Reports}
\bvolume{10}(\bissue{1}),
\bfpage{21965}
(\byear{2020}).
\doiurl{10.1038/s41598-020-78811-3}
\end{barticle}
\endbibitem

\bibitem{singhPhysiologicFacetCapsule2019}
\begin{barticle}
\bauthor{\bsnm{Singh}, \binits{S.}},
\bauthor{\bsnm{Kartha}, \binits{S.}},
\bauthor{\bsnm{Bulka}, \binits{B.A.}},
\bauthor{\bsnm{Stiansen}, \binits{N.S.}},
\bauthor{\bsnm{Winkelstein}, \binits{B.A.}}:
\batitle{Physiologic facet capsule stretch can induce pain \& upregulate matrix
  metalloproteinase-3 in the dorsal root ganglia when preceded by a
  physiological mechanical or nonpainful chemical exposure}.
\bjtitle{Clinical Biomechanics}
\bvolume{64},
\bfpage{122}--\blpage{130}
(\byear{2019}).
\doiurl{10.1016/j.clinbiomech.2018.01.009}
\end{barticle}
\endbibitem

\bibitem{cohenEpidemiologyDiagnosisTreatment2015}
\begin{barticle}
\bauthor{\bsnm{Cohen}, \binits{S.P.}}:
\batitle{Epidemiology, diagnosis, and treatment of neck pain}.
\bjtitle{Mayo Clinic Proceedings}
\bvolume{90}(\bissue{2}),
\bfpage{284}--\blpage{299}
(\byear{2015}).
\doiurl{10.1016/j.mayocp.2014.09.008}
\end{barticle}
\endbibitem

\bibitem{coteAnnualIncidenceCourse2004}
\begin{barticle}
\bauthor{\bsnm{C{\^o}t{\'e}}, \binits{P.}},
\bauthor{\bsnm{Cassidy}, \binits{D.J.}},
\bauthor{\bsnm{Carroll}, \binits{L.J.}},
\bauthor{\bsnm{Kristman}, \binits{V.}}:
\batitle{The annual incidence and course of neck pain in the general
  population: A population-based cohort study:}.
\bjtitle{Pain}
\bvolume{112}(\bissue{3}),
\bfpage{267}--\blpage{273}
(\byear{2004}).
\doiurl{10.1016/j.pain.2004.09.004}
\end{barticle}
\endbibitem

\bibitem{hoyEpidemiologyNeckPain2010}
\begin{barticle}
\bauthor{\bsnm{Hoy}, \binits{D.G.}},
\bauthor{\bsnm{Protani}, \binits{M.}},
\bauthor{\bsnm{De}, \binits{R.}},
\bauthor{\bsnm{Buchbinder}, \binits{R.}}:
\batitle{The epidemiology of neck pain}.
\bjtitle{Best Practice and Research Clinical Rheumatology}
\bvolume{24}(\bissue{6}),
\bfpage{783}--\blpage{792}
(\byear{2010}).
\doiurl{10.1016/j.berh.2011.01.019}
\end{barticle}
\endbibitem

\bibitem{banCollagenOrganizationFacet2017}
\begin{barticle}
\bauthor{\bsnm{Ban}, \binits{E.}},
\bauthor{\bsnm{Zhang}, \binits{S.}},
\bauthor{\bsnm{Zarei}, \binits{V.}},
\bauthor{\bsnm{Barocas}, \binits{V.H.}},
\bauthor{\bsnm{Winkelstein}, \binits{B.A.}},
\bauthor{\bsnm{Picu}, \binits{C.R.}}:
\batitle{Collagen organization in facet capsular ligaments varies with spinal
  region and with ligament deformation}.
\bjtitle{Journal of Biomechanical Engineering}
\bvolume{139}(\bissue{7}),
\bfpage{071009}
(\byear{2017}).
\doiurl{10.1115/1.4036019}
\end{barticle}
\endbibitem

\bibitem{zareiTissueLoadingMicrostructure2017}
\begin{barticle}
\bauthor{\bsnm{Zarei}, \binits{V.}},
\bauthor{\bsnm{Zhang}, \binits{S.}},
\bauthor{\bsnm{Winkelstein}, \binits{B.A.}},
\bauthor{\bsnm{Barocas}, \binits{V.H.}}:
\batitle{Tissue loading and microstructure regulate the deformation of embedded
  nerve fibres: Predictions from single-scale and multiscale simulations}.
\bjtitle{Journal of The Royal Society Interface}
\bvolume{14}(\bissue{135}),
\bfpage{20170326}
(\byear{2017}).
\doiurl{10.1098/rsif.2017.0326}
\end{barticle}
\endbibitem

\bibitem{hatami-marbiniScalingNonaffineDeformation2008}
\begin{barticle}
\bauthor{\bsnm{{Hatami-Marbini}}, \binits{H.}},
\bauthor{\bsnm{Picu}, \binits{R.C.}}:
\batitle{Scaling of nonaffine deformation in random semiflexible fiber
  networks}.
\bjtitle{Physical Review E}
\bvolume{77}(\bissue{6}),
\bfpage{062103}
(\byear{2008}).
\doiurl{10.1103/PhysRevE.77.062103}
\end{barticle}
\endbibitem

\bibitem{huismanInternalStressesNormal2011}
\begin{barticle}
\bauthor{\bsnm{Huisman}, \binits{E.M.}},
\bauthor{\bsnm{Lubensky}, \binits{T.C.}}:
\batitle{Internal stresses, normal modes, and nonaffinity in three-dimensional
  biopolymer networks}.
\bjtitle{Physical Review Letters}
\bvolume{106}(\bissue{8}),
\bfpage{088301}
(\byear{2011}).
\doiurl{10.1103/PhysRevLett.106.088301}
\end{barticle}
\endbibitem

\bibitem{picuMechanicsRandomFiber2011}
\begin{barticle}
\bauthor{\bsnm{Picu}, \binits{R.C.}}:
\batitle{Mechanics of random fiber networks\textemdash a review}.
\bjtitle{Soft Matter}
\bvolume{7}(\bissue{15}),
\bfpage{6768}--\blpage{6785}
(\byear{2011}).
\doiurl{10.1039/C1SM05022B}
\end{barticle}
\endbibitem

\bibitem{chandranAffineNonaffineFibril2006}
\begin{barticle}
\bauthor{\bsnm{Chandran}, \binits{P.L.}},
\bauthor{\bsnm{Barocas}, \binits{V.H.}}:
\batitle{Affine versus non-affine fibril kinematics in collagen networks:
  Theoretical studies of network behavior}.
\bjtitle{Journal of biomechanical engineering}
\bvolume{128}(\bissue{2}),
\bfpage{259}--\blpage{270}
(\byear{2006})
\end{barticle}
\endbibitem

\bibitem{shermanMaterialsScienceCollagen2015}
\begin{barticle}
\bauthor{\bsnm{Sherman}, \binits{V.R.}},
\bauthor{\bsnm{Yang}, \binits{W.}},
\bauthor{\bsnm{Meyers}, \binits{M.A.}}:
\batitle{The materials science of collagen}.
\bjtitle{Journal of the Mechanical Behavior of Biomedical Materials}
\bvolume{52},
\bfpage{22}--\blpage{50}
(\byear{2015}).
\doiurl{10.1016/j.jmbbm.2015.05.023}
\end{barticle}
\endbibitem

\bibitem{grekasCellsExploitPhase2019}
\begin{botherref}
\oauthor{\bsnm{Grekas}, \binits{G.}},
\oauthor{\bsnm{Proestaki}, \binits{M.}},
\oauthor{\bsnm{Rosakis}, \binits{P.}},
\oauthor{\bsnm{Notbohm}, \binits{J.}},
\oauthor{\bsnm{Makridakis}, \binits{C.}},
\oauthor{\bsnm{Ravichandran}, \binits{G.}}:
Cells exploit a phase transition to establish interconnections in fibrous
  extracellular matrices.
arXiv:1905.11246 [cond-mat, physics:physics]
(2019)
{\href{https://arxiv.org/abs/1905.11246}{{arxiv:1905.11246}}}
{[cond-mat, physics:physics]}
\end{botherref}
\endbibitem

\bibitem{hatami-marbiniMultiscaleModelingSemiflexible2013}
\begin{barticle}
\bauthor{\bsnm{{Hatami-Marbini}}, \binits{H.}},
\bauthor{\bsnm{Shahsavari}, \binits{A.}},
\bauthor{\bsnm{Picu}, \binits{R.C.}}:
\batitle{Multiscale modeling of semiflexible random fibrous structures}.
\bjtitle{Computer-Aided Design}
\bvolume{45}(\bissue{1}),
\bfpage{77}--\blpage{83}
(\byear{2013}).
\doiurl{10.1016/j.cad.2011.10.002}
\end{barticle}
\endbibitem

\bibitem{picuPoissonContractionFiber2018}
\begin{barticle}
\bauthor{\bsnm{Picu}, \binits{R.C.}},
\bauthor{\bsnm{Deogekar}, \binits{S.}},
\bauthor{\bsnm{Islam}, \binits{M.R.}}:
\batitle{Poisson's contraction and fiber kinematics in tissue: Insight from
  collagen network simulations}.
\bjtitle{Journal of Biomechanical Engineering}
\bvolume{140}(\bissue{2}),
\bfpage{021002}
(\byear{2018}).
\doiurl{10.1115/1.4038428}
\end{barticle}
\endbibitem

\bibitem{licupStressControlsMechanics2015}
\begin{barticle}
\bauthor{\bsnm{Licup}, \binits{A.J.}},
\bauthor{\bsnm{M{\"u}nster}, \binits{S.}},
\bauthor{\bsnm{Sharma}, \binits{A.}},
\bauthor{\bsnm{Sheinman}, \binits{M.}},
\bauthor{\bsnm{Jawerth}, \binits{L.M.}},
\bauthor{\bsnm{Fabry}, \binits{B.}},
\bauthor{\bsnm{Weitz}, \binits{D.A.}},
\bauthor{\bsnm{MacKintosh}, \binits{F.C.}}:
\batitle{Stress controls the mechanics of collagen networks}.
\bjtitle{Proceedings of the National Academy of Sciences}
\bvolume{112}(\bissue{31}),
\bfpage{9573}--\blpage{9578}
(\byear{2015}).
\doiurl{10.1073/pnas.1504258112}
\end{barticle}
\endbibitem

\bibitem{headMechanicalResponseSemiflexible2005}
\begin{barticle}
\bauthor{\bsnm{Head}, \binits{D.A.}},
\bauthor{\bsnm{Levine}, \binits{A.J.}},
\bauthor{\bsnm{MacKintosh}, \binits{F.C.}}:
\batitle{Mechanical response of semiflexible networks to localized
  perturbations}.
\bjtitle{Physical Review E}
\bvolume{72}(\bissue{6}),
\bfpage{061914}
(\byear{2005}).
\doiurl{10.1103/PhysRevE.72.061914}
\end{barticle}
\endbibitem

\bibitem{levineDeformationFieldSemiflexible2004}
\begin{barticle}
\bauthor{\bsnm{Levine}, \binits{A.J.}},
\bauthor{\bsnm{Head}, \binits{D.A.}},
\bauthor{\bsnm{MacKintosh}, \binits{F.C.}}:
\batitle{The deformation field in semiflexible networks}.
\bjtitle{Journal of Physics: Condensed Matter}
\bvolume{16}(\bissue{22}),
\bfpage{2079}--\blpage{2088}
(\byear{2004}).
\doiurl{10.1088/0953-8984/16/22/006}
\end{barticle}
\endbibitem

\bibitem{chandranDeterministicMaterialBasedAveraging2007}
\begin{barticle}
\bauthor{\bsnm{Chandran}, \binits{P.L.}},
\bauthor{\bsnm{Barocas}, \binits{V.H.}}:
\batitle{Deterministic material-based averaging theory model of collagen gel
  micromechanics}.
\bjtitle{Journal of Biomechanical Engineering}
\bvolume{129}(\bissue{2}),
\bfpage{137}--\blpage{147}
(\byear{2007}).
\doiurl{10.1115/1.2472369}
\end{barticle}
\endbibitem

\bibitem{stylianopoulosVolumeaveragingTheoryStudy2007}
\begin{barticle}
\bauthor{\bsnm{Stylianopoulos}, \binits{T.}},
\bauthor{\bsnm{Barocas}, \binits{V.H.}}:
\batitle{Volume-averaging theory for the study of the mechanics of collagen
  networks}.
\bjtitle{Computer Methods in Applied Mechanics and Engineering}
\bvolume{196}(\bissue{31-32}),
\bfpage{2981}--\blpage{2990}
(\byear{2007}).
\doiurl{10.1016/j.cma.2006.06.019}
\end{barticle}
\endbibitem

\bibitem{aghvamiMultiscaleMechanicalSimulations2013}
\begin{barticle}
\bauthor{\bsnm{Aghvami}, \binits{M.}},
\bauthor{\bsnm{Barocas}, \binits{V.H.}},
\bauthor{\bsnm{Sander}, \binits{E.A.}}:
\batitle{Multiscale mechanical simulations of cell compacted collagen gels}.
\bjtitle{Journal of Biomechanical Engineering}
\bvolume{135}(\bissue{7}),
\bfpage{071004}
(\byear{2013}).
\doiurl{10.1115/1.4024460}
\end{barticle}
\endbibitem

\bibitem{stylianopoulosMultiscaleStructurebasedModeling2007}
\begin{barticle}
\bauthor{\bsnm{Stylianopoulos}, \binits{T.}},
\bauthor{\bsnm{Barocas}, \binits{V.H.}}:
\batitle{Multiscale, structure-based modeling for the elastic mechanical
  behavior of arterial walls}.
\bjtitle{Journal of Biomechanical Engineering}
\bvolume{129}(\bissue{4}),
\bfpage{611}--\blpage{618}
(\byear{2007}).
\doiurl{10.1115/1.2746387}
\end{barticle}
\endbibitem

\bibitem{zareiImagebasedMultiscaleMechanical2017}
\begin{barticle}
\bauthor{\bsnm{Zarei}, \binits{V.}},
\bauthor{\bsnm{Liu}, \binits{C.J.}},
\bauthor{\bsnm{Claeson}, \binits{A.A.}},
\bauthor{\bsnm{Akkin}, \binits{T.}},
\bauthor{\bsnm{Barocas}, \binits{V.H.}}:
\batitle{Image-based multiscale mechanical modeling shows the importance of
  structural heterogeneity in the human lumbar facet capsular ligament}.
\bjtitle{Biomechanics and Modeling in Mechanobiology}
\bvolume{16}(\bissue{4}),
\bfpage{1425}--\blpage{1438}
(\byear{2017}).
\doiurl{10.1007/s10237-017-0896-4}
\end{barticle}
\endbibitem

\bibitem{chanImagebasedMultiscaleMechanical2019}
\begin{barticle}
\bauthor{\bsnm{Chan}, \binits{V.W.L.}},
\bauthor{\bsnm{Tobin}, \binits{W.R.}},
\bauthor{\bsnm{Zhang}, \binits{S.}},
\bauthor{\bsnm{Winkelstein}, \binits{B.A.}},
\bauthor{\bsnm{Barocas}, \binits{V.H.}},
\bauthor{\bsnm{Shephard}, \binits{M.S.}},
\bauthor{\bsnm{Picu}, \binits{C.R.}}:
\batitle{Image-based multi-scale mechanical analysis of strain amplification in
  neurons embedded in collagen gel}.
\bjtitle{Computer Methods in Biomechanics and Biomedical Engineering}
\bvolume{22}(\bissue{2}),
\bfpage{113}--\blpage{129}
(\byear{2019}).
\doiurl{10.1080/10255842.2018.1538414}
\end{barticle}
\endbibitem

\bibitem{zagarTwoFundamentalMechanisms2015}
\begin{barticle}
\bauthor{\bsnm{{\v Z}agar}, \binits{G.}},
\bauthor{\bsnm{Onck}, \binits{P.R.}},
\bauthor{\bsnm{{van~der~Giessen}}, \binits{E.}}:
\batitle{Two fundamental mechanisms govern the stiffening of cross-linked
  networks}.
\bjtitle{Biophysical Journal}
\bvolume{108}(\bissue{6}),
\bfpage{1470}--\blpage{1479}
(\byear{2015}).
\doiurl{10.1016/j.bpj.2015.02.015}
\end{barticle}
\endbibitem

\end{thebibliography}

\end{document}